\documentclass[11pt,a4paper]{article}

\usepackage{jcappub}
\usepackage{graphicx}
\usepackage{xcolor}
\usepackage[caption=false]{subfig}
\usepackage{mathrsfs,mathtools}
\usepackage{physics,amssymb}
\usepackage{bm}
\usepackage{braket}
\usepackage{listings}
\usepackage{cases}
\usepackage{comment}
\usepackage{soul}
\usepackage{cancel}
\usepackage{cases}
\usepackage[utf8]{inputenc}
\usepackage{url}
\usepackage{longtable}
\usepackage[normalem]{ulem}
\usepackage{xspace}
\hypersetup{colorlinks=true
,urlcolor=DARKBLUE
,anchorcolor=DARKBLUE
,citecolor=DARKBLUE
,filecolor=DARKBLUE
,linkcolor=DARKBLUE
,menucolor=DARKBLUE
,linktocpage=true
,pdfproducer=medialab
}

\newcommand{\ee}{\mathrm{e}}

\newcommand{\Mpl}{M_\mathrm{Pl}}

\newcommand{\uth}{\mathrm{th}}
\newcommand{\fNL}{f_\mathrm{NL}}
\newcommand{\fNLmin}{f_\mathrm{NL,min}}
\newcommand{\pk}{\mathrm{pk}}
\newcommand{\MS}{\mathrm{MS}}
\newcommand{\ren}{\mathrm{ren}}
\newcommand{\PBH}{\mathrm{PBH}}
\newcommand{\DM}{\mathrm{DM}}
\newcommand{\uNG}{\mathrm{NG}}
\newcommand{\mono}{\mathrm{mono}}
\newcommand{\uII}{\mathrm{II}}
\newcommand{\tot}{\mathrm{tot}}
\newcommand{\eq}{\mathrm{eq}}

\newcommand{\calC}{\mathcal{C}}

\newcommand{\uF}{\mathrm{F}}
\newcommand{\calF}{\mathcal{F}}

\newcommand{\uG}{\mathrm{G}}

\newcommand{\bfk}{\mathbf{k}}

\newcommand{\um}{\mathrm{m}}

\newcommand{\calO}{\mathcal{O}}

\newcommand{\calP}{\mathcal{P}}

\newcommand{\bfx}{\mathbf{x}}
\newcommand{\uW}{\mathrm{W}}

\newcommand{\bae}[1]{\begin{align} #1 \end{align}}

\newcommand{\bfe}[4]{
\begin{figure} 
	\centering
	\includegraphics[#1]{#2}
	\caption{#3}
	\label{#4}
\end{figure}}
\newcommand{\beae}[1]{\begin{equation}\begin{aligned} #1 \end{aligned}\end{equation}}

\definecolor{MONZA}{HTML}{CF000F}
\definecolor{DARKBLUE}{HTML}{00008b}
\definecolor{DARKMAGENTA}{HTML}{8b008b}
\definecolor{DARKCYAN}{HTML}{008B8B}
\definecolor{DARKORANGE}{HTML}{FF8C00}

\title{Primordial black holes in peak theory with a non-Gaussian tail}

\author[a,b]{Naoya Kitajima,}
\author[c,d]{Yuichiro Tada,}
\author[e,f]{Shuichiro Yokoyama,}
\author[d]{and Chul-Moon Yoo}

\affiliation[a]{Frontier Research Institute for Interdisciplinary Sciences, Tohoku University, \\
Aramaki aza Aoba 6-3, Aobaku, Sendai 980-8578 Japan}
\affiliation[b]{Department of Physics, Tohoku University, \\
Aramaki aza Aoba 6-3, Aobaku, Sendai 980-8578 Japan}
\affiliation[c]{Institute for Advanced Research, Nagoya University, \\
Furocho Chikusaku Nagoya, Aichi 464-8601 Japan}
\affiliation[d]{Department of Physics, Nagoya University, \\
Furocho Chikusaku Nagoya, Aichi 464-8602 Japan}
\affiliation[e]{Kobayashi Maskawa Institute, Nagoya University, \\
Chikusa, Aichi 464-8602, Japan}
\affiliation[f]{Kavli IPMU (WPI), UTIAS, The University of Tokyo, \\
Kashiwa, Chiba 277-8583, Japan}

\emailAdd{naoya.kitajima.c2@tohoku.ac.jp}
\emailAdd{tada.yuichiro@e.mbox.nagoya-u.ac.jp}
\emailAdd{shu@kmi.nagoya-u.ac.jp}
\emailAdd{yoo@gravity.phys.nagoya-u.ac.jp}

\abstract{In this paper, we update the peak theory for the estimation of the primordial black hole (PBH) abundance, 
particularly by implementing the critical behavior in the estimation of the PBH mass and employing the averaged compaction function for the PBH formation criterion to relax the profile dependence.
We apply our peak theory to a specific non-Gaussian feature called the \emph{exponential tail}, which is characteristic in ultra slow-roll models of inflation. With this type of non-Gaussianity, the probability of a large perturbation is not suppressed by the Gaussian factor but decays only exponentially, so the PBH abundance is expected to be much enhanced.
Not only do we confirm this enhancement even compared to the case of the corresponding nonlinearity parameter $\fNL=5/2$, but also we find that the resultant PBH mass spectrum has a characteristic maximal mass which is not seen in the simple Press--Schechter approach.
}
\arxivnumber{2109.00791}

\begin{document}

\begin{flushright}
TU-1130
\end{flushright}

\maketitle
\flushbottom

\section{Introduction}

The primordial black hole has been attracting more and more attention recently.
Contrary to the ordinary stellar-origin black hole, it has been hypothesized that black holes can be produced even in the early radiation-dominated universe by the gravitational collapse of extraordinarily overdense regions, dubbed \emph{primordial black hole} (PBH)~\cite{Hawking:1971ei,Carr:1974nx,Carr:1975qj}.
Not only is it an exciting candidate of the unknown dark matter (DM) (see Ref.~\cite{Carr:2020gox,Carr:2016drx} for a recent review), but the PBH has been motivated in many intriguing cosmological/astrophysical scenarios: PBHs as seeds of supermassive black holes in galaxies (see, e.g., Ref.~\cite{Carr:2018rid}), as the possible Planet 9 of our solar system~\cite{Scholtz:2019csj,Witten:2020ifl}, as the gravitational lensing objects inferred by the Optical Gravitational Lensing Experiment~\cite{Niikura:2019kqi}, as a trigger of $r$-process by the neutron-star capture~\cite{Fuller:2017uyd}, as a source of the fast radio burst~\cite{Fuller:2017uyd,Kainulainen:2021rbg}, etc.
Among these motivations, the recently most important feature is its profound relation to the gravitational wave (GW).
PBHs can form binaries due to gravitation by themselves during the radiation-dominated era after their formations, and explain the merger GWs observed by the LIGO--Virgo collaboration if they constitute subpercent fraction of the total dark matter (see the review~\cite{Sasaki:2018dmp}).
Furthermore, the large primordial perturbations necessary to sizable PBHs can also cause stochastic GWs by themselves through the second-order interaction between the scalar and tensor metric perturbations~\cite{Saito:2008jc,Saito:2009jt}.
Such stochastic GWs can be detected in the near future: ground-based interferometers represented by Einstein Telescope~\cite{ET} corresponds to $\sim10^{13}\,\mathrm{g}$ PBHs, space-based interferometers such as LISA~\cite{2017arXiv170200786A} (as well as Taiji~\cite{Ruan:2018tsw}, TianQin~\cite{TianQin:2015yph}, and DECIGO~\cite{Kawamura:2011zz}) 
to $\sim10^{20}\,\mathrm{g}$ (the best point as a DM candidate), and the pulsar timing array (PTA) which will be updated by SKA~\cite{Janssen:2014dka} to $\sim1M_\odot$ (possibly related to merger GWs), as well-known detection techniques, 
and also several novel attempts may bridge the gap between SKA and LISA (e.g., Refs.~\cite{Bustamante-Rosell:2021daj,Blas:2021mpc,Blas:2021mqw}) and between LISA and ET (e.g., Refs.~\cite{AEDGE:2019nxb,Badurina:2019hst}). 
In fact, the NANOGrav collaboration, one current PTA observation, has recently reported a possible detection of such stochastic GWs and attracted attention ~\cite{NANOGrav:2020bcs}. 
In order to utilize these stochastic GWs as a ``cross-check" of PBHs, it is necessary to correctly relate the PBH abundance and the GW amplitude, which requires the accurate estimation of the PBH abundance from given primordial perturbations (see, e.g., Refs.~\cite{Ando:2018qdb,DeLuca:2019qsy,Young:2019osy,Young:2020xmk} for attempts at the accurate estimation of the PBH abundance, Refs.~\cite{Bugaev:2008gw,Josan:2009qn,Sato-Polito:2019hws,Kalaja:2019uju,Gow:2020bzo,Unal:2020mts,Kimura:2021sqz} for power spectrum reconstruction from the PBH abundance, and Refs.~\cite{Kohri:2018awv,Cai:2018dig,Unal:2018yaa,Inomata:2018epa} for relations between the primordial perturbations and the induced GWs). 

On the other primordial perturbation side, a specific non-Gaussian feature called \emph{exponential tail} is drawing attention in the context of, e.g., the stochastic inflation~\cite{Pattison:2017mbe,Ezquiaga:2019ftu,Figueroa:2020jkf,Pattison:2021oen}. In the standard inflationary scenario, the primordial curvature perturbation $\zeta$ is assumed to follow the Gaussian distribution and thus the probability density of large $\zeta$ is suppressed by the Gaussian factor $\propto\exp\pqty{-\frac{\zeta^2}{2\sigma_0^2}}$ with the variance $\sigma_0^2=\braket{\zeta^2}$.
However, in the ultra slow-roll case (or extremely flat potential case), it is reported that the large $\zeta$ probability is suppressed only by the exponential factor $\propto\exp(-\Lambda\zeta)$ with some decaying factor $\Lambda$. Therefore, it can be much amplified compared to the Gaussian case.
It is understood as that, in the ultra slow-roll case, the probability of extraordinarily long-lasting inflation due to the inflaton fluctuation is not much suppressed just like a bridge between the standard small perturbation inflation and the so-called eternal inflation.
The PBH abundance is determined by the large perturbation probability rather than the properties of small perturbations, and in fact, it has been pointed out that 
even a small non-Gaussian correction by the $\fNL$ term can strongly affect the PBH abundance (e.g., Refs.~\cite{Bullock:1996at,Ivanov:1997ia,Yokoyama:1998pt,  Hidalgo:2007vk,Byrnes:2012yx,Bugaev:2013vba,Young:2015cyn,Nakama:2016gzw,Ando:2017veq,Franciolini:2018vbk,Cai:2018dig,Atal:2018neu,Passaglia:2018ixg,Taoso:2021uvl}). 
The effect of the exponential tail will be much significant than this $\fNL$ correction.
Also from the practical aspect that PBHs are often realized in ultra slow-roll models, this non-Gaussian correction is important (see, e.g., Refs.~\cite{Atal:2019cdz,Atal:2019erb,Biagetti:2021eep} for works discussing the exponential tail effect on the PBH abundance).

In this paper, we update the peak theory for the PBH abundance as one of the currently most accurate schemes and apply it to the exponential tail distribution, in a series of previous works~\cite{Yoo:2018kvb,Yoo:2019pma,Yoo:2020dkz}.
Our peak theory i) seeks the peaks of the Laplacian of the curvature perturbation, $-\Delta\zeta$, which is free from the contamination by very long-wavelength curvature perturbations, ii) can take account of the local-type non-Gaussian correction including the exponential tail, iii) adopts the scaling behavior for a more realistic PBH mass function, and iv) employs the averaged compaction function as the threshold to relax the profile dependence particularly in non-Gaussian cases.
We also compare it with the commonly used Press--Schechter approach in Appendix~\ref{sec: comparison to other approaches} in detail.
In the exponential tail case, we found not only the PBH abundance is much more amplified than the $\fNL$ expansion as expected, but also it results in a specific PBH mass spectrum with a hard cut on a maximal mass which is not seen in the Press--Schechter approach.

The paper is organized as follows. In Sec.~\ref{sec: peak theory}, we formulate the algorithm of the peak theory in a general context including the local-type non-Gaussianity. In Sec.~\ref{sec: monochromatic curvature perturbations}, we assume a monochromatic peak power spectrum~\eqref{eq: monochromatic power} for the curvature perturbation for simplicity and exemplify our peak theory in the Gaussian (Sec.~\ref{sec: Gaussian}), the $\fNL$-type non-Gaussian (Sec.~\ref{sec: fNL}), and the exponential tail cases (Sec.~\ref{sec: exponential tail}).
Sec.~\ref{sec: discussion and conclusions} is devoted to the discussion and conclusions. The computation in the Press--Schechter approach is reviewed in Appendix~\ref{sec: comparison to other approaches} as a comparison to our peak theory.

\section{Peak theory with a local-type non-Gaussian correction}\label{sec: peak theory}

In this section, we formulate the peak theory for the PBH abundance in a general setup, including a local-type non-Gaussian correction.

\subsection{Algorithm overview}

The abundance of PBHs has been so far investigated in various approaches.
The very first and simplest methods are represented by the so-called Press--Schechter formalism~\cite{Press:1973iz}.
In essence, this formalism estimates the PBH abundance by the probability that the locally-averaged value of some random field (such as the density contrast) exceeds a given threshold value.
The original work by Carr~\cite{Carr:1975qj} shows that the threshold value in terms of the density contrast is evaluated as $\delta_\uth\sim w=1/3$ in the radiation-dominated era, where the equation-of-state parameter $w=p/\rho$ is the ratio of the pressure $p$ to the energy density $\rho$ of the cosmic fluid.
Subsequently, many analytic or numerical works have updated this threshold value in various ways~(see, e.g., Refs.~\cite{Shibata:1999zs,Musco:2004ak,Harada:2013epa}).
This variety of the threshold value however implies the difficulty in the Press--Schechter approach: the PBH formation criterion is not a simple matter of the universal threshold for a one-point value of some averaged random field, but we have to deal with the detailed local profile (i.e., the spatial dependence) of overdensities~\cite{Nakama:2013ica,Escriva:2019nsa,Atal:2019erb}.

While the PBH formation condition depending on the profile needs independent numerical/analytical work, the statistical property of the peak profile itself can be dealt with in the \emph{peak theory}.
According to the original work by Bardeen, Bond, Kaiser, and Szalay~\cite{Bardeen:1985tr} (see also Refs.~\cite{Yoo:2018kvb,Yoo:2019pma,Yoo:2020dkz}), 
the high peak of some random Gaussian field $g(\bfx)$ typically takes the spherically symmetric profile (here and in what follows, the hat denotes typical-profile-related quantities), 
\bae{\label{eq: ghat}
    \hat{g}(r)=\mu_0\bqty{\frac{1}{1-\gamma_1^2}\pqty{\psi_0(r)+\frac{1}{3}R_1^2\Delta\psi_0(r)}-k_1^2\frac{1}{\gamma_1(1-\gamma_1^2)}\frac{\sigma_0}{\sigma_2}\pqty{\gamma_1^2\psi_0(r)+\frac{1}{3}R_1^2\Delta\psi_0(r)}},
}
with two random variables $\eval{\mu_0=g}_{r=0}$ and $\eval{k_1^2=-\Delta g}_{r=0}/\mu_0$ representing the height and width of the peak, and the statistical parameters and (generalized) two point functions of $g$,
\beae{\label{eq: stat params}
    &\sigma_n^2=\int\frac{\dd{k}}{k}k^{2n}\calP_g(k), &
    &\psi_n(r)=\frac{1}{\sigma_n^2}\int\frac{\dd{k}}{k}k^{2n}\frac{\sin(kr)}{kr}\calP_g(k), \\
    &\gamma_n=\frac{\sigma_n^2}{\sigma_{n-1}\sigma_{n+1}}, &
    &R_n=\frac{\sqrt{3}\sigma_n}{\sigma_{n+1}} \qc \text{for odd $n$,}
}
determined by $g$'s power spectrum
\bae{
    \calP_g(k)=\frac{k^3}{2\pi^2}\int\dd[3]{\bfx}\ee^{-i\bfk\cdot\bfx}\Braket{g\pqty{\frac{\bfx}{2}}g\pqty{-\frac{\bfx}{2}}}.
}
$\Delta$ is the Laplacian. The radius center is at the extremal point: $\eval{\bm{\nabla}g=\mathbf{0}}_{r=0}$.
The number density of such a peak in a comoving volume (i.e., the comoving number density of positive extremal points of $g$ in a high peak limit) is furthermore expected statistically as
\bae{\label{eq: npk mu0 k1}
    n_\pk^{(\mu_0,k_1)}\dd{\mu_0}\dd{k_1}=\frac{2\cdot3^{3/2}}{(2\pi)^{3/2}}\mu_0k_1\frac{\sigma_2^2}{\sigma_0\sigma_1^3}f\pqty{\frac{\mu_0k_1^2}{\sigma_2}}P_1^{(1)}\pqty{\frac{\mu_0}{\sigma_0},\frac{\mu_0k_1^2}{\sigma_2}}\dd{\mu_0}\dd{k_1},
}
where
\bae{
    f(\xi)&=\frac{1}{2}\xi(\xi^2-3)\pqty{\erf\bqty{\frac{1}{2}\sqrt{\frac{5}{2}}\xi}+\erf\bqty{\sqrt{\frac{5}{2}}\xi}} \nonumber \\
    &\quad +\sqrt{\frac{2}{5\pi}}\Bqty{\pqty{\frac{8}{5}+\frac{31}{4}\xi^2}\exp\bqty{-\frac{5}{8}\xi^2}+\pqty{-\frac{8}{5}+\frac{1}{2}\xi^2}\exp\bqty{-\frac{5}{2}\xi^2}}, \label{eq: f}
}
and
\bae{
    P_1^{(n)}(\nu,\xi)=\frac{1}{2\pi\sqrt{1-\gamma_n^2}}\exp\bqty{-\frac{1}{2}\pqty{\nu^2+\frac{(\xi-\gamma\nu)^2}{1-\gamma_n^2}}}.
}
Here $\erf(z)$ denotes the error function $\erf(z)=\frac{2}{\sqrt{\pi}}\int^z_0\ee^{-t^2}\dd{t}$.
Consequently, once the PBH formation criterion is clarified for the typical profile~\eqref{eq: ghat} in terms of $\mu_0$ and $k_1$, in principle the PBH abundance can be strictly determined within the range of the high peak approximation (if the source field follows the Gaussian distribution, though).

On the other PBH formation condition side, several works suggest using the so-called \emph{compaction function}~\cite{Shibata:1999zs,Harada:2015yda}.
In a spherically symmetric case, the compaction function $\calC$ is defined by the difference between the Misner--Sharp mass $M_\MS$ and the expected mass $M_\uF$ in the background universe within the same areal radius $R$, which is given on a comoving slice by\footnote{Our compaction function is twice larger than the definition in Refs.~\cite{Yoo:2018kvb,Yoo:2019pma,Yoo:2020dkz} so that it is equivalent to the smoothed density perturbation at the horizon reentry~\eqref{eq: calC and delta}.}
\bae{
    \calC=\frac{M_\MS-M_\uF}{4\pi\Mpl^2R} \qc
    M_\MS=4\pi\int^R_0\rho{\tilde{R}}^2\dd{\tilde{R}} \qc
    M_\uF=\frac{4\pi}{3}\bar{\rho}R^3.
}
$\bar{\rho}$ is the background energy density and $\Mpl$ is the reduced Planck mass.
Note that it can be understood as the average of the comoving density contrast $\delta=(\rho-\bar{\rho})/\bar{\rho}$ at the horizon reentry $RH=1$ as
\bae{\label{eq: calC and delta}
    \calC=(RH)^2\times\left.\pqty{4\pi\int^R_0\delta{\tilde{R}}^2\dd{\tilde{R}}}\middle/\pqty{\frac{4\pi}{3}R^3},\right.
}
making use of the Friedmann equation $3\Mpl^2H^2=\bar{\rho}$.
Defining the comoving curvature perturbation $\zeta$ via the spatial metric by\footnote{Note that the sign in $\zeta$'s definition is opposite to Refs.~\cite{Yoo:2018kvb,Yoo:2019pma,Yoo:2020dkz} so that positive values of our $\zeta$ correspond to overdensities as commonly used in the cosmological context.}
\bae{
    \dd{s_3^2}=a^2(t)\ee^{2\zeta(t,\bfx)}\tilde{\gamma}_{ij}\dd{x^i}\dd{x^j} \qc
    \det\tilde{\gamma}=1,
}
the comoving density contrast is related to $\zeta$ on a superhorizon scale as
\bae{
    \delta=-\frac{4(1+w)}{5+3w}\frac{1}{a^2H^2}\ee^{-5\zeta/2}\Delta\ee^{\zeta/2}\quad\overset{w=1/3}{=}-\frac{8}{9}\frac{1}{a^2H^2}\ee^{-5\zeta/2}\Delta\ee^{\zeta/2}.
}
Making use of the areal radius expression $R(r)=a\ee^{\zeta}r$, one summarizes the compaction function in a simple form:
\bae{\label{eq: compaction C}
    \calC(r)=\frac{2}{3}\bqty{1-(1+r\zeta^\prime)^2}.
}
It is useful as it is conserved on a superhorizon scale contrary to the density contrast itself. 

Letting $r_\um$ correspond to the innermost maximum of $\calC(r)$, the associated areal radius $R(r_\um)$ is understood as the proper ``size" of the overdensity, and if its maximum $\calC_\um=\calC(r_\um)$ exceeds some threshold, it is assumed to be a PBH soon after its horizon reentry $R(r_\um)H=1$. Note that $R(r)$ can be non-monotonic in $r$ for a large perturbation (i.e., large $\mu$) and such a perturbation is called \emph{type II}~\cite{Kopp:2010sh}. In this case, the maximum compaction $\calC_\um$ can decrease for a large perturbation and be smaller than the threshold, so that a too large overdensity apparently does not form a PBH.
However, Ref.~\cite{Kopp:2010sh} has shown that a type II perturbation always leads to a PBH formation irrespectively of the value of $\calC_\um$.
Unfortunately, the mass of the resultant PBH has not been clarified well.
In this paper, we hence neglect the type II PBH as it is probabilistically suppressed anyway, and consider only type I perturbations where $R(r)$ is a monotonic function.
For a type I perturbation, the maximum $\calC_\um$ is monotonically increasing with respect to the perturbation amplitude $\mu$, and therefore the threshold condition on $\calC_\um$ can be converted to that on $\mu$ without any subtlety.
Note that the first derivative of the compaction function is given by
\bae{
    \calC^\prime(r)=-\frac{4}{3}\pqty{1+r\zeta^\prime}\pqty{\zeta^\prime+r\zeta^{\prime\prime}},
}
while the monotonicity of $R(r)=a\ee^{\zeta}r$ for type I perturbations ensures
\bae{
    \dv{R}{r}=a\ee^\zeta\pqty{1+r\zeta^\prime}>0.
}
Therefore, the extremal condition of the compaction function can be reduced to
\bae{\label{eq: extremal condition}
    \pqty{\zeta^\prime+r\zeta^{\prime\prime}}_{r=r_\um}=0.
}

Although the compaction function is a useful indicator, actually the threshold value in terms of $\calC_\um$ slightly depends on the peak profile~\cite{Escriva:2019phb,Atal:2019erb}.
The more universal and profile-independent threshold is found on the average of the compaction function:
\bae{\label{eq: barCm}
    \bar{\calC}_\um=\left.\pqty{4\pi\int_0^{R(r_\um)}\calC(r)\tilde{R}^2(r)\dd{\tilde{R}(r)}}\middle/\pqty{\frac{4\pi}{3}R^3(r_\um)}. \right.
}
In this work, we adopt the threshold value $\bar{\calC}_\uth=2/5$ shown in the literature~\cite{Escriva:2019phb,Atal:2019erb}.
Note that this averaged $\bar{\calC}_\um$ can decrease for large $\mu$ even in a type I range as we will see below. However, the compaction $\calC_\um$ itself is increasing even in that case and thus a PBH is assumed to be formed definitely.
Namely, the threshold for $\calC_\um$ slightly depending on the profile is obtained for each $k$ by the smallest $\mu$ with which the averaged compaction $\bar{\calC}_\um$ exceeds a universal threshold $\bar{\calC}_\uth$.
In other words, we will follow the procedure that we
first find the threshold $\mu_\uth$ in terms of the perturbation amplitude as the minimum $\mu$ satisfying $\bar{\calC}_\um\geq\bar{\calC}_\uth$ for each $k$, and then uniformly assume the perturbation with $\mu>\mu_\uth$ to be a PBH, regardless of $\bar{\calC}_\um$.

Let us then choose a proper source Gaussian field $g$ for PBH formation.
As the curvature perturbation $\zeta$ is a superhorizon-conserved variable, the peak of $\zeta$ can be contaminated by the much longer-wavelength perturbation though such an offset of the curvature perturbation should not affect the local physics like the PBH formation but be merely renormalized into the scale factor $a$ (see, e.g., Ref.~\cite{Young:2014ana}).
In Ref.~\cite{Yoo:2020dkz}, it is therefore suggested to use the Laplacian $\zeta_2\coloneqq-\Delta\zeta$ as the source field $g$ instead of $\zeta$ itself.
Defining the statistical parameters~\eqref{eq: stat params} with respect to $\calP_\zeta$ rather than $\calP_{\Delta\zeta}$ for convenience, the typical profile for Gaussian $\zeta_2$ is obtained by the increment of indices $n\to n+2$ from Eq.~\eqref{eq: ghat} as
\bae{
    \hat{\zeta}_2(r)&=\hat{\zeta}^\uG_2(r) \nonumber \\
    &=\mu_2\bqty{\frac{1}{1-\gamma_3^2}\pqty{\psi_2(r)+\frac{1}{3}R_3^2\Delta\psi_2(r)}-k_3^2\frac{1}{\gamma_3(1-\gamma_3^2)}\frac{\sigma_2}{\sigma_4}\pqty{\gamma_3^2\psi_2(r)+\frac{1}{3}R_3^2\Delta\psi_2(r)}},
}
with $\eval{\mu_2=-\Delta\zeta}_{r=0}$ and $\eval{k_3^2=\Delta\Delta\zeta}_{r=0}/\mu_2$. 
Noting that the two point functions are the eigen series of the Laplacian as
\bae{
    \Delta\psi_n(r)=-\frac{\sigma_{n+1}^2}{\sigma_n^2}\psi_{n+1}(r),
}
one can integrate $\hat{\zeta}^\uG_2$ with the regularity condition $\eval{\partial_r\hat{\zeta}^\uG}_{r=0}=0$ as
\bae{\label{eq: zetabar G}
    \hat{\zeta}(r)&=\hat{\zeta}^\uG(r) \nonumber \\
    &=\tilde{\mu}_2\bqty{\frac{1}{1-\gamma_3^2}\pqty{\psi_1(r)+\frac{1}{3}R_3^2\Delta\psi_1(r)}-\frac{\tilde{k}_3^2}{\gamma_3(1-\gamma_3^2)}\pqty{\gamma_3^2\psi_1(r)+\frac{1}{3}R_3^2\Delta\psi_1(r)}}+\zeta^\uG_\infty,
}
with an integration constant $\eval{\zeta^\uG_{\infty}=\hat{\zeta}^\uG}_{r\to\infty}$ as a new random variable.
Here $\tilde{\mu}_2=\frac{\sigma_1^2}{\sigma_2^2}\mu_2$ and $\tilde{k}_3=\sqrt{\frac{\sigma_2}{\sigma_4}}k_3$ are dimensionless parameters.
Ref.~\cite{Yoo:2020dkz} shows that $\zeta^\uG_\infty$ also follows the Gaussian distribution with the probability
\bae{
    p(\zeta^\uG_\infty)=\pqty{\frac{1-\gamma_3^2}{2\pi D\sigma_0^2}}^{1/2}\exp\bqty{-\frac{1-\gamma_3^2}{2D\sigma_0^2}{\zeta^\uG_\infty}^2},
}
for fixed $\tilde{\mu}_2$ and $\tilde{k}_3$, where
\bae{
    D=1-\gamma_1^2-\gamma_2^2-\gamma_3^2+2\gamma_1\gamma_2\gamma_3 \qc
    \gamma_2=\frac{\sigma_2^2}{\sigma_0\sigma_4}.
}
However, as mentioned, $\zeta^\uG_\infty$ is an irrelevant offset, so that we use the renormalized $\zeta$,
\bae{
    \hat{\zeta}_\ren=\hat{\zeta}^\uG_\ren(r)=\hat{\zeta}^\uG(r)-\zeta^\uG_\infty,
}
by redefining the scale factor as $a\to\tilde{a}=a\ee^{\zeta^\uG_\infty}$ (we will omit the tilde of the renormalized scale factor hereafter).
Though the compaction function~\eqref{eq: compaction C} is not affected by the constant $\zeta^\uG_\infty$ anyway, this renormalization is necessary for a precise PBH mass evaluation which we describe below.

We can now estimate the PBH number density in terms of $\tilde{\mu}_2$ and $\tilde{k}_3$, but it is better to be expressed in the PBH mass as a mass function.
It is widely accepted that the resultant PBH mass for a peak around the threshold follows a scaling relation
\bae{\label{eq: MPBH}
    M(\tilde{\mu}_2,\tilde{k}_3)=K(\tilde{k}_3)\pqty{\tilde{\mu}_2-\tilde{\mu}_{2,\uth}(\tilde{k}_3)}^\gamma M_H(\tilde{\mu}_2,\tilde{k}_3),
}
with a universal power $\gamma\simeq0.36$~\cite{Choptuik:1992jv,Evans:1994pj,Koike:1995jm,Niemeyer:1997mt,Niemeyer:1999ak,Hawke:2002rf,Musco:2008hv}.\footnote{This scaling relation is sometimes written in terms of the density contrast as $M\propto(\delta-\delta_\uth)^\gamma$ in the literature. However the density contrast is not a monotonic function of the amplitude of the perturbation as we mentioned right after Eq.~\eqref{eq: barCm}. In this paper, we instead choose $\tilde{\mu}_2$ monotonic in the perturbation amplitude as a proper scaling parameter.}
$K(\tilde{k}_3)$ is a profile-depending order-unity parameter. As its precise profile dependence has not been fully clarified (see, e.g., Ref.~\cite{Escriva:2019nsa} for a relevant work), we uniformly approximate it as $K\simeq1$ for simplicity.
$M_H$ is the horizon mass at the horizon reentry of the maximal radius, $\hat{R}_\um H=a\ee^{\hat{\zeta}_\ren(\hat{r}_\um)}r_\um H=1$, where $\hat{r}_\um$ corresponds to the maximum of the typical compaction $\hat{\calC}(r)=\frac{2}{3}\bqty{1-(1+r\hat{\zeta}^\prime)^2}$. 
$\tilde{\mu}_{2,\uth}(\tilde{k}_3)$ is the PBH threshold in terms of $\tilde{\mu}_2$, which can be converted from the threshold condition $\bar{\calC}_\um(\tilde{\mu}_{2,\uth},\tilde{k}_3)=\bar{\calC}_\uth$.
It should be noted that not only is this scaling relation unapplicable to type II perturbations, but also it starts to fail for too large type I perturbations.
However, in a realistic case, probabilistically the PBH abundance is mainly contributed by small perturbations, and therefore we uniformly use this scaling relation, neglecting this subtlety.
Once this relation is accepted, one can change the independent variables from $(\tilde{\mu}_2,\tilde{k}_3)$ to, e.g., $(\ln M,\tilde{k}_3)$, and the PBH comoving number density with the mass in the range of $[M,M\ee^{\dd{\ln M}}]$ is obtained as
\bae{\label{eq: nPBH}
    n_\PBH(M)\dd{\ln M}=\pqty{\int_{\tilde{\mu}_2(M,\tilde{k}_3)\geq\tilde{\mu}_{2,\uth}(\tilde{k}_3)}n_\pk^{(\tilde{\mu}_2,\tilde{k}_3)}\pqty{\tilde{\mu}_2(M,\tilde{k}_3),\tilde{k}_3}\abs{\dv{\ln M}{\tilde{\mu}_2}}^{-1}\dd{\tilde{k}_3}}\dd{\ln M},
}
from the peak number density
\bae{
    n_\pk^{(\mu_2,k_3)}\dd{\mu_2}\dd{k_3}&=n_\pk^{(\tilde{\mu}_2,\tilde{k}_3)}\dd{\tilde{\mu}_2}\dd{\tilde{k}_3} \nonumber \\
    &=\frac{2\cdot3^{3/2}}{(2\pi)^{3/2}}\frac{\sigma_2^2\sigma_4^3}{\sigma_1^4\sigma_3^3}\tilde{\mu}_2\tilde{k}_3f\pqty{\frac{\sigma_2}{\sigma_1^2}\tilde{\mu}_2\tilde{k}_3^2}P_1^{(3)}\pqty{\frac{\sigma_2}{\sigma_1^2}\tilde{\mu}_2,\frac{\sigma_2}{\sigma_1^2}\tilde{\mu}_2\tilde{k}_3^2}\dd{\tilde{\mu}_2}\dd{\tilde{k}_3},
}
where $n_\pk^{(\mu_2,k_3)}$ itself can be found by the indices increment of $n_\pk^{(\mu_0,k_1)}$~\eqref{eq: npk mu0 k1}.
As the comoving number density of PBHs is conserved from their formation to today, neglecting their merger, accretion, evaporation, etc., the current energy density ratio $f_\PBH$ of PBHs to total dark matters is directly calculated by
\bae{\label{eq: fPBH}
    f_\PBH(M) \dd{\ln M} =\frac{Mn_\PBH(M)}{3\Mpl^2H_0^2\Omega_\DM} \dd{\ln M},
}
with the current Hubble parameter $H_0$ and the density parameter for total dark matter $\Omega_\DM=\rho_\DM/(3\Mpl^2H_0^2)$.

\subsection{Local-type non-Gaussianity}\label{sec: local-type NG}

The above discussion is valid only if the source curvature perturbation follows the Gaussian distribution.
In general, there is no simple generalization of our approach to an arbitrary probability distribution.
However, if the full statistics is well approximated by the so-called local-type non-Gaussianity, i.e., the full field $\zeta(\bfx)$ is given by some function of a Gaussian kernel $\zeta^\uG(\bfx)$ at the same spatial point as $\zeta(\bfx)=\calF_\uNG(\zeta^\uG(\bfx))$, one may take some correction into account as proposed in Ref.~\cite{Yoo:2019pma}.
It relies on one assumption that \emph{the peak position of the full field $\zeta$ corresponds to that of the Gaussian kernel $\zeta^\uG$}.
If it holds, the typical profile of the full field is given by that of the Gaussian kernel as $\hat{\zeta}(r)=\calF_\uNG(\hat{\zeta}^\uG(r))$ where $\hat{\zeta}^\uG(r)$ is given by Eq.~\eqref{eq: zetabar G}, and then the procedure in the previous subsection can be applied.
For the consistency of this assumption, it is required that the non-Gaussianity is small enough so that the function $\calF_\uNG(z)$ is monotonically increasing (i.e., $\calF^\prime_\uNG(z)>0$) at least around the threshold condition.
We also note that the offset $\zeta^\uG_\infty$ has a physical meaning in this case.
The redefinition of the scale factor can kill only the offset of the full field, so that the renormalized curvature perturbation is given by
\bae{\label{eq: zetaren}
    \hat{\zeta}_\ren(r)=\calF_\uNG(\hat{\zeta}^\uG(r))-\calF_\uNG(\zeta^\uG_\infty),
}
which in general depends on the offset $\zeta^\uG_\infty$ of the Gaussian kernel for a nonlinear function $\calF_\uNG$.
Therefore, the independent random parameters are now three as $(\tilde{\mu}_2,\tilde{k}_3,\zeta^\uG_\infty)$.
Though the PBH abundance is in principle calculable even in this case, practically the inverse function of the PBH mass~\eqref{eq: MPBH} is hard to be found in general, which is necessary to the integration for $n_\PBH$~\eqref{eq: nPBH}.
In the $\fNL$ expansion approach $\calF_\uNG(z)\simeq z+\frac{3}{5}\fNL z^2$, one can invert $M(\tilde{\mu}_2,\tilde{k}_3,\zeta^\uG_\infty)$ but we address this issue in a separated paper~\cite{KTYY2}.
Note that, in the case where the power spectrum of the curvature perturbation has a sharp enough peak, the offset $\zeta_\infty^\uG$ is restricted to zero and thus the renormalization~\eqref{eq: zetaren} does not matter as we see in the next section.

\section{Monochromatic curvature perturbations}\label{sec: monochromatic curvature perturbations}

In this section, we focus on the monochromatic curvature perturbation power spectrum,
\bae{\label{eq: monochromatic power}
    \calP_{\zeta^\uG}(k)=A_{\zeta^\uG}\delta(\ln k-\ln k_*),
}
for simplicity, and see several example PBH mass spectra, including local-type non-Gaussian corrections.
This monochromatic limit significantly simplifies the peak theory.
The variances are obtained as $\sigma_n^2\to A_{\zeta^\uG} k_*^{2n}$ and the all $\gamma$ parameters asymptote to unity as $\gamma_{1,2,3}\to1$.
The random parameters $\tilde{k}_3$ and $\zeta^\uG_\infty$ are then restricted to $1$ and $0$, respectively, as\footnote{One can first introduce a regulator $\sigma_{\ln k}$ as $\calP_{\zeta^\uG}(k)=\frac{A_{\zeta^\uG}}{\sqrt{2\pi\sigma_{\ln k}^2}}\exp\pqty{-\frac{(\ln k-\ln k_*)^2}{2\sigma_{\ln k}^2}}$ and take the limit $\sigma_{\ln k}\to0$ to obtain them. \label{footnote: regulator}}
\bae{
    P_1^{(3)}\pqty{\frac{\sigma_2}{\sigma_1}^2\tilde{\mu}_2,\frac{\sigma_2}{\sigma_1^2}\tilde{\mu}_2\tilde{k}_3^2}\to\frac{A_{\zeta^\uG}}{2\tilde{\mu}_2}\frac{1}{\sqrt{2\pi A_{\zeta^\uG}}}\ee^{-\tilde{\mu}_2^2/(2A_{\zeta^\uG})}\delta(\tilde{k}_3-1) \qc
    p(\zeta^\uG_\infty)\to\delta(\zeta^\uG_\infty).
}
Therefore, the independent parameters are effectively reduced to only $\tilde{\mu}_2$.\footnote{In this limit, our approach is equivalent to the prescription in Refs.~\cite{Atal:2019cdz,Atal:2019erb}.}

\subsection{Gaussian}\label{sec: Gaussian}

Let us begin with the case where $\zeta$ is exactly Gaussian.
First note that the typical profile~\eqref{eq: zetabar G} is simplified as 
\bae{
    \eval{\hat{\zeta}^\uG(r)}_{\tilde{k}_3=\sqrt{\gamma_3}}=\tilde{\mu}_2\psi_1(r)+\zeta^\uG_\infty,
}
for $\tilde{k}_3=\sqrt{\gamma_3}$ which holds in the monochromatic case.
Therefore, with the restriction of $\tilde{k}_3=1$ and $\zeta^\uG_\infty=0$, we can focus on the following simple profile,\footnote{This is again derived strictly by introducing the regulator $\sigma_{\ln k}$ similarly to footnote~\ref{footnote: regulator}.}
\bae{
    \hat{\zeta}(r)=\hat{\zeta}^\uG_\mono(r)=\tilde{\mu}_2\frac{\sin x}{x},
}
with the renormalized radius $x=k_*r$.
Let us then consider the type I condition.
One can numerically find that the areal radius $\hat{R}(r)=\frac{a}{k_*}x\exp\pqty{\tilde{\mu}_2\frac{\sin x}{x}}$ is monotonic (i.e., $\hat{R}^\prime(r)$ is always positive for arbitrary $r$) if $\tilde{\mu}_2<\tilde{\mu}_{2,\uII}\simeq0.941$.
We therefore restrict $\tilde{\mu}_2$ up to this value $\tilde{\mu}_{2,\uII}$ ad hoc.

\begin{figure}
    \centering
    \begin{tabular}{c}
        \begin{minipage}{0.486\hsize}
            \centering
            \includegraphics[width=0.95\hsize]{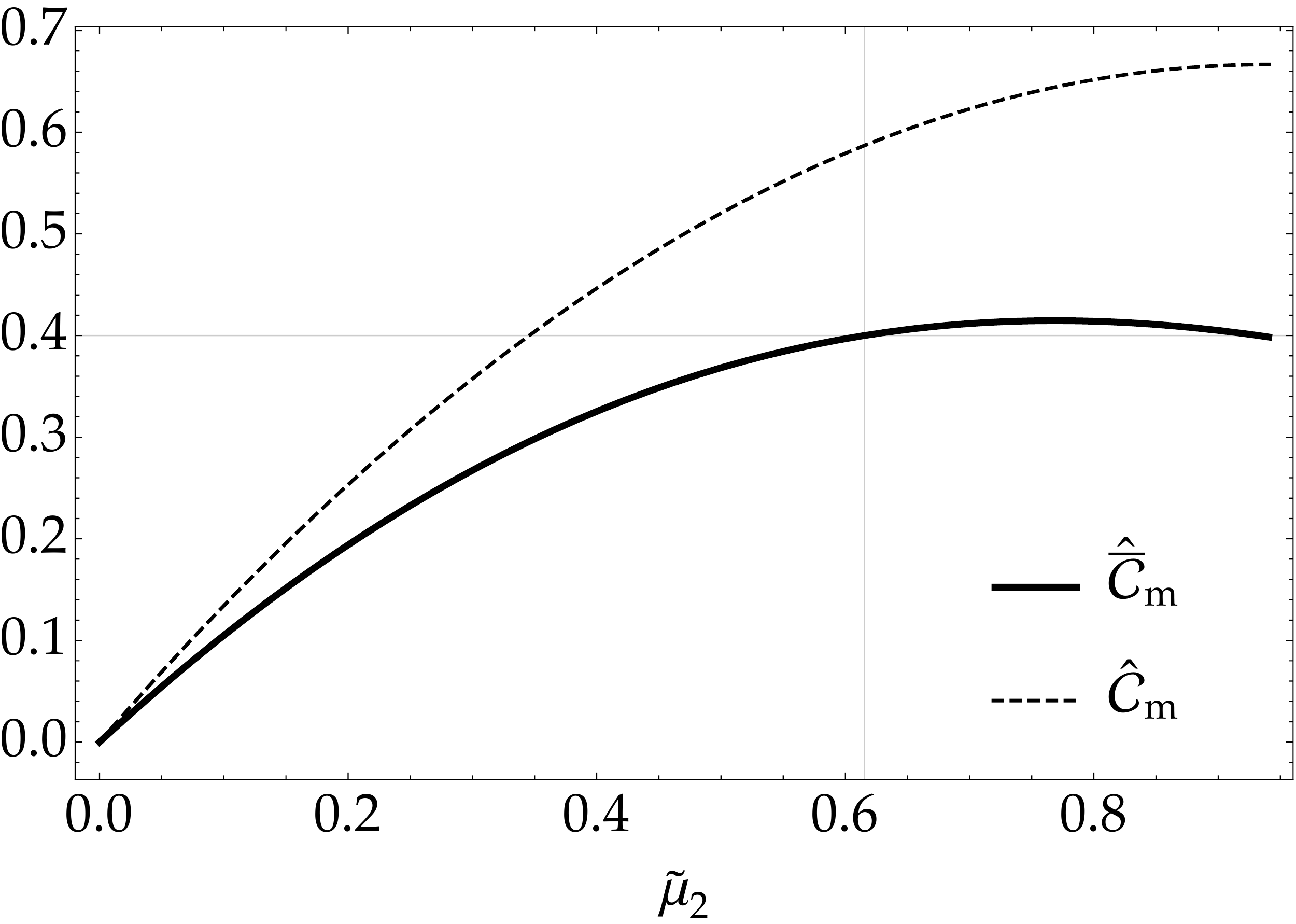}
        \end{minipage}
        \begin{minipage}{0.514\hsize}
            \centering
            \includegraphics[width=0.95\hsize]{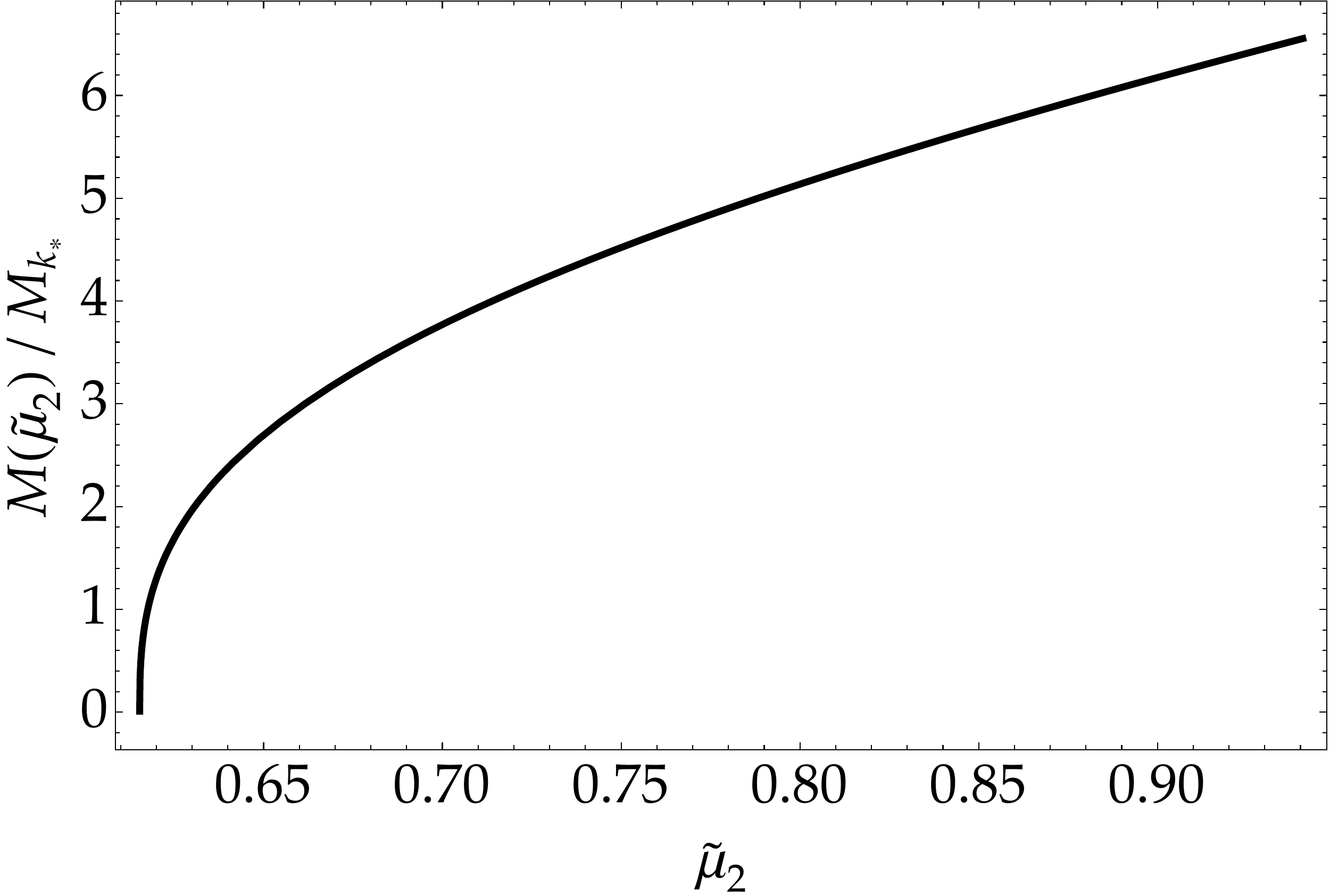}
        \end{minipage}
    \end{tabular}
    \caption{\emph{Left}: the averaged compaction function $\hat{\bar{\calC}}_\um$~\eqref{eq: barCm} (thick) with respect to $\tilde{\mu}_2$. The horizontal thin line indicates the threshold $\bar{\calC}_\uth=2/5$, while the vertical one is the corresponding threshold in terms of $\tilde{\mu}_2$: $\tilde{\mu}_{2,\uth}\simeq0.615$.
    As a comparison, we also show the compaction maximum $\hat{\calC}_\um$ itself in the dashed line. This is indeed monotonically increasing in $\tilde{\mu}_2$ for type I perturbations $<\tilde{\mu}_{2,\uII}\simeq0.941$.
    \emph{Right}: the PBH mass~\eqref{eq: PBH mass Gauss} with respect to $\tilde{\mu}_2$ normalized by the $k_*$-related horizon mass $M_{k_*}=M_k(k_*)$.}
    \label{fig: Cm and M Gauss}
\end{figure}

The extremal condition of the compaction function is given by Eq.~\eqref{eq: extremal condition} as
\bae{
    0=\pqty{\hat{\zeta}^\prime+r\hat{\zeta}^{\prime\prime}}_{r=\hat{r}_\um}=-k_*\tilde{\mu}_2\frac{\hat{x}_\um\cos\hat{x}_\um+(\hat{x}_\um^2-1)\sin\hat{x}_\um}{\hat{x}_\um^2},
}
which can be numerically solved as $\hat{x}_\um\simeq2.74$ irrespectively of $\tilde{\mu}_2$.
Making use of the expression of the typical compaction function
\bae{
    \hat{\calC}(r)=\frac{2}{3}\bqty{1-\pqty{1+r\hat{\zeta}^\prime}^2}=\frac{2}{3}\bqty{1-\pqty{1+\tilde{\mu}_2\pqty{\cos x-\frac{\sin x}{x}}}^2},
}
the mean compaction $\hat{\bar{\calC}}_\um$~\eqref{eq: barCm} can also be computed numerically as a function of $\tilde{\mu}_2$.
The left panel of Fig.~\ref{fig: Cm and M Gauss} shows the result in comparison to the threshold $\bar{\calC}_\uth=2/5$.
We interpret the smaller crossing point $\tilde{\mu}_{2,\uth}\simeq0.615$ as the threshold in $\tilde{\mu}_2$ and assume that larger $\tilde{\mu}_2$ always forms a PBH even if apparently $\hat{\bar{\calC}}_\um<\bar{\calC}_\uth$.

\begin{figure}
    \centering
    \begin{tabular}{c}
        \begin{minipage}{0.5\hsize}
            \centering
            \includegraphics[width=0.95\hsize]{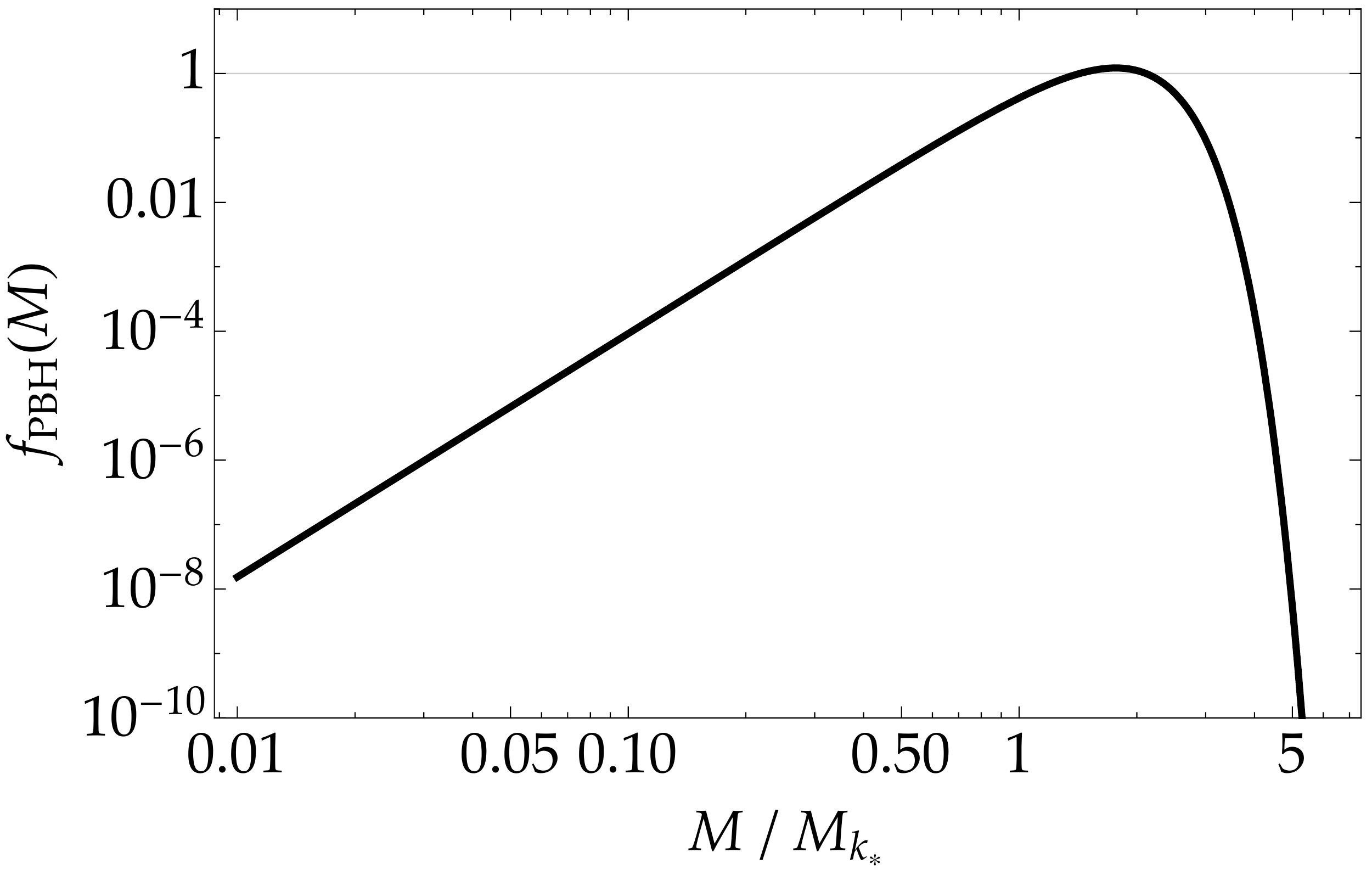}
        \end{minipage}
        \begin{minipage}{0.5\hsize}
            \centering
            \includegraphics[width=0.95\hsize]{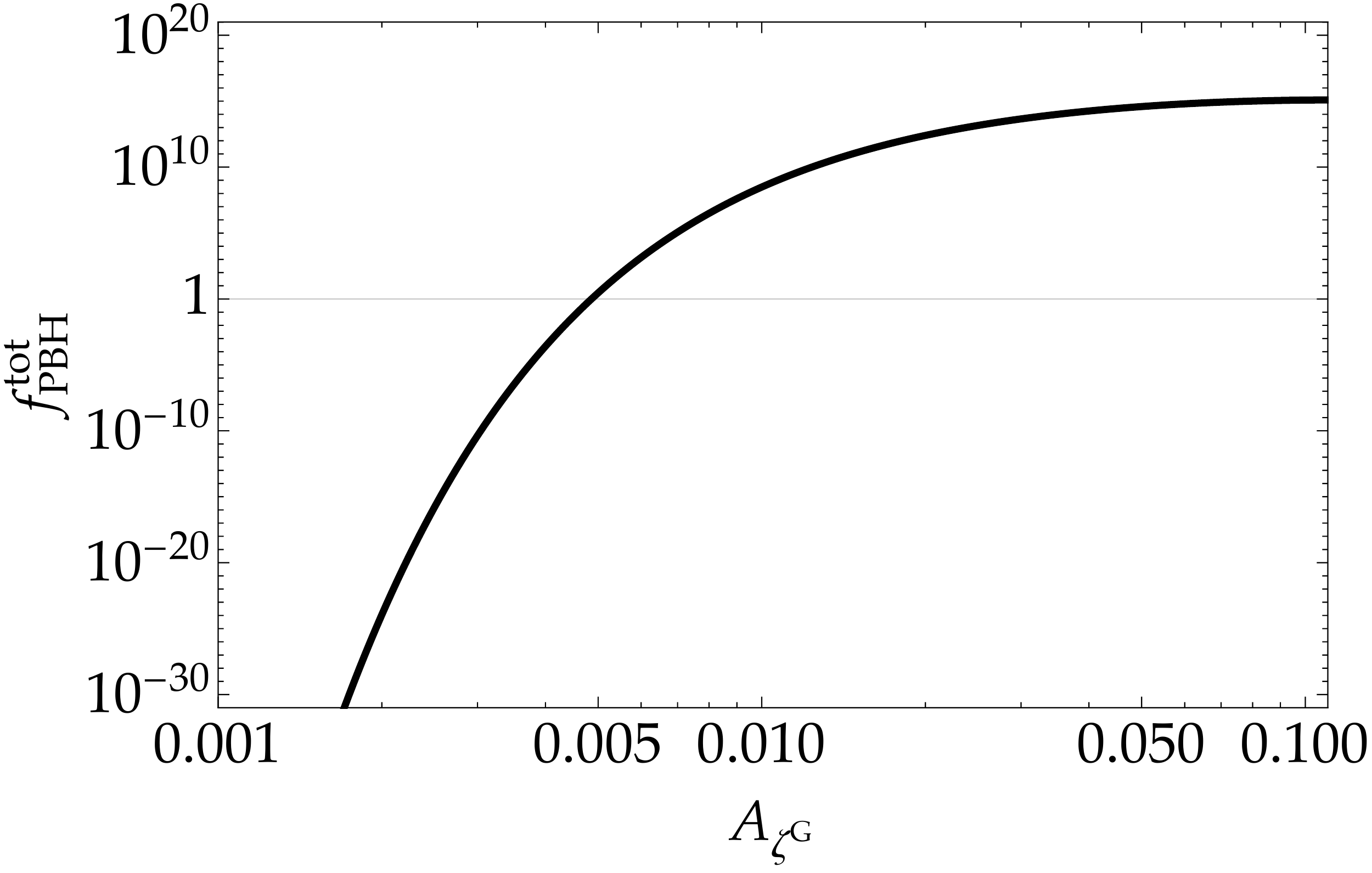}
        \end{minipage}
    \end{tabular}
    \caption{\emph{Left}: the PBH mass spectrum~\eqref{eq: fNL Gauss} for $k_*=1.56\times10^{13}\,\mathrm{Mpc}^{-1}$ and $A_{\zeta^\uG}=4.863\times10^{-3}$ power spectrum of the curvature perturbation. \emph{Right}: the total PBH abundance~\eqref{eq: fPBH tot Gauss} as a function of the curvature perturbation amplitude $A_{\zeta^\uG}$.}
    \label{fig: fPBH Gauss}
\end{figure}

The PBH mass is approximated by the critical behavior~\eqref{eq: MPBH}.
It is useful to first introduce the $k$-related horizon mass $M_k(k)$, i.e., the horizon mass at the horizon reentry of the comoving wavenumber $k$: $aH/k=1$.
It is computed as (see, e.g., Ref.~\cite{Tada:2019amh})
\bae{
    M_k(k)\simeq10^{20}\pqty{\frac{g_*}{106.75}}^{-1/6}\pqty{\frac{k}{1.56\times10^{13}\,\mathrm{Mpc}^{-1}}}^{-2}\,\mathrm{g},
}
where $g_*$ denotes the effective degrees of freedom for the energy density of the cosmic fluid at $k$'s horizon reentry. We hereafter uniformly assume $g_*=106.75$ for our relevant PBH mass range.
Recalling the horizon reentry condition of the maximal radius, $a\ee^{\hat{\zeta}(\hat{r}_\um)}\hat{r}_\um H=1$, one finds that the corresponding horizon mass $M_H$ is given by $M_k(\hat{r}_\um^{-1}\ee^{-\hat{\zeta}_\um})$ with $\hat{\zeta}_\um=\hat{\zeta}(\hat{r}_\um)$. 
Therefore the PBH mass is expressed as
\bae{\label{eq: PBH mass Gauss}
    M(\tilde{\mu}_2)=\hat{x}_\um^2\ee^{2\hat{\zeta}_\um(\tilde{\mu}_2)}K\pqty{\tilde{\mu}_2-\tilde{\mu}_{2,\uth}}^\gamma M_k(k_*),
}
which is exhibited in the right panel of Fig.~\ref{fig: Cm and M Gauss}.
The Jacobian then reads
\bae{\label{eq: dlnMdmu}
    \abs{\dv{\ln M}{\tilde{\mu}_2}}=\abs{2\hat{\zeta}_\um^\prime(\tilde{\mu}_2)+\frac{\gamma}{\tilde{\mu}_2-\tilde{\mu}_{2,\uth}}}\simeq0.282+\frac{0.36}{\tilde{\mu}_2-\tilde{\mu}_{2,\uth}}.
}
The PBH number density~\eqref{eq: nPBH} can be calculated with use of this Jacobian by
\bae{
    n_\PBH(M)\dd{\ln M}=n_\pk^{(\tilde{\mu}_2)}(\tilde{\mu}_2(M))\abs{\dv{\ln M}{\tilde{\mu}_2}}^{-1}\dd{\ln M},
}
with the peak number density
\bae{
    n_\pk^{(\tilde{\mu}_2)}(\tilde{\mu}_2)\dd{\tilde{\mu}_2}=\int n_\pk^{(\tilde{\mu}_2,\tilde{k}_3)}(\tilde{\mu}_2,\tilde{k}_3)\dd{\tilde{k}_3}=\pqty{\frac{3}{2\pi}}^{3/2}k_*^3f\pqty{\frac{\tilde{\mu}_2}{\sqrt{A_{\zeta^\uG}}}}\frac{1}{\sqrt{2\pi A_{\zeta^\uG}}}\ee^{-\tilde{\mu}_2^2/(2A_{\zeta^\uG})}\dd{\tilde{\mu}_2}.
}
One finally obtains the current PBH abundance~\eqref{eq: fPBH} as
\bae{
    f_\PBH(M)=\pqty{\frac{\Omega_\DM h^2}{0.12}}^{-1}\pqty{\frac{M}{10^{20}\,\mathrm{g}}}\pqty{\frac{k_*}{1.56\times10^{13}\,\mathrm{Mpc}^{-1}}}^3 \nonumber \\
    \times\pqty{\frac{\abs{\dv{\ln M}{\tilde{\mu}_2}}^{-1}f\pqty{\frac{\tilde{\mu}_2(M)}{\sqrt{A_{\zeta^\uG}}}}P_\uG\pqty{\tilde{\mu}_2(M),A_{\zeta^\uG}}}{5.3\times10^{-16}}}, \label{eq: fNL Gauss}
}
with $f(\xi)$~\eqref{eq: f} and the Gaussian distribution $P_\uG(x,\sigma^2)=\frac{1}{\sqrt{2\pi\sigma^2}}\ee^{-x^2/(2\sigma^2)}$.
The left panel of Fig.~\ref{fig: fPBH Gauss} shows an example $f_\PBH(M)$ for $k_*=1.56\times10^{13}\,\mathrm{Mpc}^{-1}$ (corresponding to $M_k(k_*)=M_{k_*}=10^{20}\,\mathrm{g}$) and $A_{\zeta^\uG}=4.863\times10^{-3}$, for which the total PBH abundance
\bae{\label{eq: fPBH tot Gauss}
    f_\PBH^\tot=\int f_\PBH(M)\dd{\ln M},
}
becomes equivalent to total dark matters: $f_\PBH^\tot=1$,
and the right panel exhibits this total abundance as a function of the amplitude of the power spectrum $A_{\zeta^\uG}$.
The low-mass tail in the left panel is a characteristics of the scaling relation~\eqref{eq: MPBH}.
Noting that $\dv{\ln f_\PBH}{\tilde{\mu}_2}$ is well approximated by $\dv{\ln M}{\tilde{\mu}_2}-\pqty{\dv{\ln M}{\tilde{\mu}_2}}^{-1}\dv[2]{\ln M}{\tilde{\mu}_2}$ around the threshold due to the divergence feature of $\dv{\ln M}{\tilde{\mu}_2}\sim\frac{\gamma}{\tilde{\mu}_2-\tilde{\mu}_{2,\uth}}$~\eqref{eq: dlnMdmu},
one finds that the power $\dv{\ln f_\PBH}{\ln M}=\pqty{\dv{\ln M}{\tilde{\mu}_2}}^{-1}\dv{\ln f_\PBH}{\tilde{\mu}_2}$ is given by $1+\gamma^{-1}\simeq3.78$ in the low-mass range (i.e., $f_\PBH\propto M^{1+\gamma^{-1}}$). This behavior is universal even in the non-Gaussian cases as one can see below.

One remark is that the PBH abundance is not simply given by the Gaussian distribution $P_\uG(\tilde{\mu}_2,A_{\zeta^\uG})$ but corrected by $f\pqty{\frac{\tilde{\mu}_2}{\sqrt{A_{\zeta^\uG}}}}$ (even aside from the critical behavior effect $\abs{\dv{\ln M}{\tilde{\mu}_2}}^{-1}$, which is in common with the Press--Schechter approach), due to the extremal condition $\bm{\nabla}\zeta_2=\mathbf{0}$.
This is one significant difference from the simple Press--Schechter approach where the PBH abundance is estimated by the probability that a Gaussian field (if the source field is Gaussian) exceeds some threshold value.
The (latest) Press--Schechter approach is reviewed in Appendix~\ref{sec: comparison to other approaches} as a comparison.

\subsection[$\fNL$ expansion]{\boldmath $\fNL$ expansion}\label{sec: fNL}

\begin{figure}
    \centering
    \begin{tabular}{c}
        \begin{minipage}{0.5\hsize}
            \centering
            \includegraphics[width=0.95\hsize]{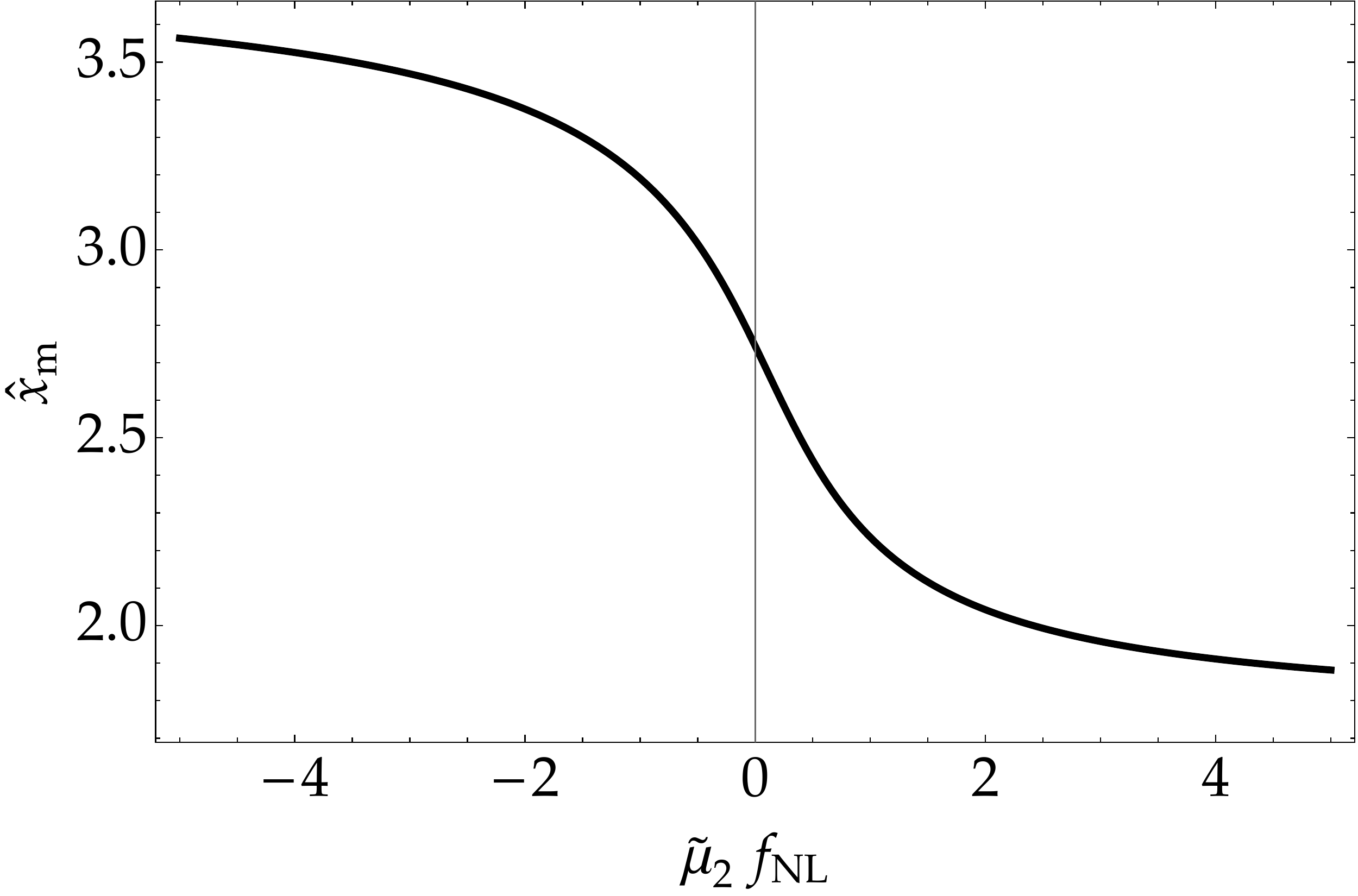}
        \end{minipage}
        \begin{minipage}{0.5\hsize}
            \centering
            \includegraphics[width=0.95\hsize]{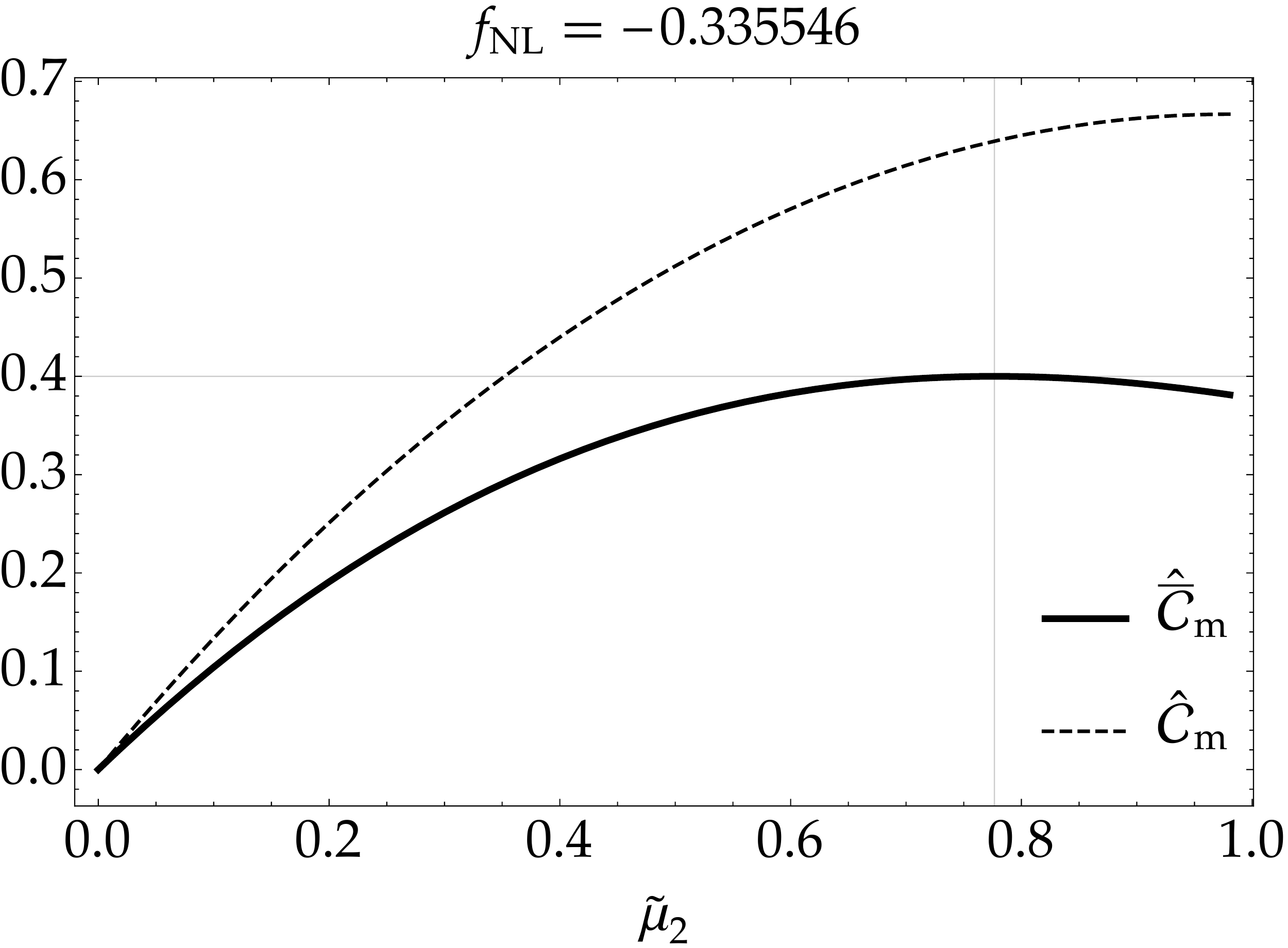}
        \end{minipage} \\
        \begin{minipage}{0.5\hsize}
            \centering
            \includegraphics[width=0.95\hsize]{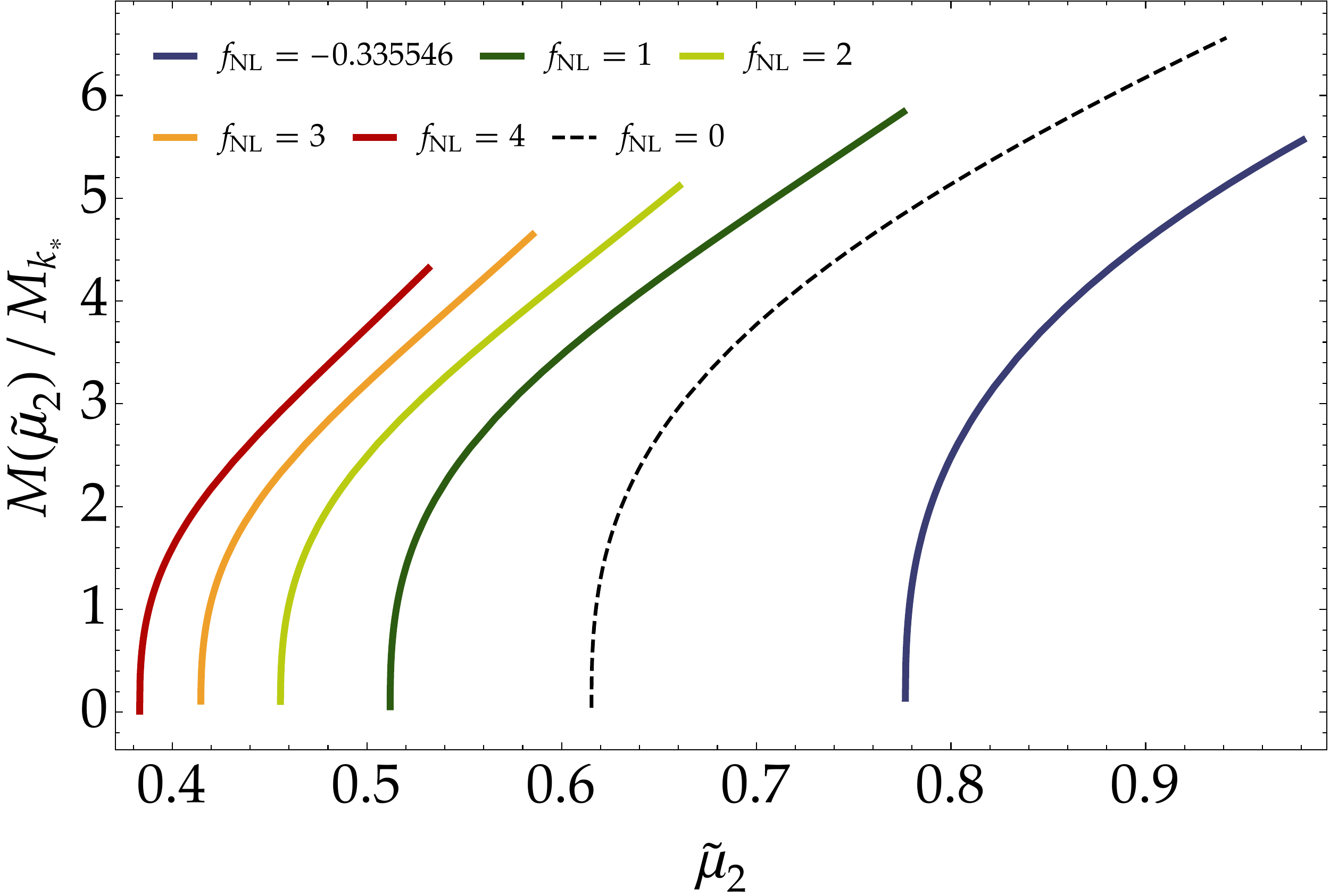}
        \end{minipage}
    \end{tabular}
    \caption{\emph{Top-Left}: the maximal radius $\hat{x}_\um$ as a funtion of the combination $\tilde{\mu}_2\fNL$.
    \emph{Top-Right}: the averaged compaction function $\hat{\bar{\calC}}_\um$ (thick) and the compaction maximum $\hat{\calC}_\um$ (dashed) for $\fNL=\fNLmin\simeq-0.336$. The horizontal line shows the threshold $\bar{\calC}_\uth=2/5$ and the vertical one indicates the corresponding threshold in $\tilde{\mu}_2$: $\tilde{\mu}_{2,\uth}(\fNLmin)\simeq0.777$.
    Though the averaged compaction reaches the threshold $\bar{\calC}_\uth=2/5$ only at its maximal point $\tilde{\mu}_{2,\uth}(\fNLmin)$ and decreases for larger $\tilde{\mu}_2$, the compaction maximum $\hat{\calC}_\um$ is monotonically increasing similarly to the Gaussian case and thus we interpret perturbations with $\tilde{\mu}_2\geq\tilde{\mu}_{2,\uth}(\fNLmin)$ to be PBHs.
    \emph{Bottom}: the PBH mass~\eqref{eq: PBH mass fNL} for $\fNL=\fNLmin\simeq-0.336$, $1$, $2$, $3$, and $4$ from right to left as well as the Gaussian case $\fNL=0$ (dashed). We only plot them for type I perturbations.}
    \label{fig: xm and M fNL}
\end{figure}

We then consider the local-type non-Gaussian correction.
In this subsection, we focus on the standard $\fNL$ expansion, i.e., the quadratic expansion of the nonlinear function $\calF_\uNG(\zeta^\uG(\bfx))$ as\footnote{Sometimes the $\fNL$ expansion is defined with the subtraction of the variance as
\bae{\label{eq: fNL expansion}
    \zeta=\zeta^\uG+\frac{3}{5}\fNL\pqty{(\zeta^\uG)^2-\braket{(\zeta^\uG)^2}},
}
to ensure the zero-mean of $\zeta$: $\braket{\zeta}=0$.
In our formulation, this constant offset is however removed anyway by the renormalization of $\zeta$~\eqref{eq: zetaren}.}
\bae{
    \zeta(\bfx)=\zeta^\uG(\bfx)+\frac{3}{5}\fNL\pqty{\zeta^\uG(\bfx)}^2,
}
where the coefficient $3/5$ is a convention. 
As described in Sec.~\ref{sec: local-type NG}, we assume that the typical profile for the full field $\zeta(\bfx)$ is simply given by the Gaussian typical profile as
\bae{\label{eq: typical zeta fNL}
    \hat{\zeta}(r)=\hat{\zeta}_\mono(r)=\hat{\zeta}^\uG_\mono(r)+\frac{3}{5}\fNL\pqty{\hat{\zeta}^\uG_\mono(r)}^2 \qc
    \hat{\zeta}^\uG_\mono(r)=\tilde{\mu}_2\frac{\sin x}{x}.
}
We need not care about the global offset because $\zeta^\uG_\infty$ is fixed to zero, i.e., the global offset automatically vanishes as $\eval{\hat{\zeta}(r)}_{r\to\infty}=0$ in the monochromatic case. $\tilde{k}_3$ is also fixed to unity.
Therefore $\tilde{\mu}_2$ is the only remaining random parameter similarly to the Gaussian case. The difference comes from the threshold condition because the compaction function is defined for the full field $\hat{\zeta}(r)$:
\bae{
    \!\!\!
    \hat{\calC}(r)=\frac{2}{3}\bqty{1-\pqty{1+r\hat{\zeta}^\prime}^2}=\frac{2}{3}\bqty{1-\Bqty{1+\tilde{\mu}_2\pqty{\cos x-\frac{\sin x}{x}}\pqty{1+\frac{6}{5}\fNL\tilde{\mu}_2\frac{\sin x}{x}}}^2}.
}
For the type I perturbation where the areal radius $\hat{R}(r)=a\ee^{\hat{\zeta}(r)}r$ is monotonic, the compaction's extremal condition reads
\bae{
    0&=\pqty{\hat{\zeta}^\prime+r\hat{\zeta}^{\prime\prime}}_{r=\hat{r}_\um} \nonumber \\
    &\propto\frac{\sin\hat{x}_\um-\hat{x}_\um(\cos\hat{x}_\um+\hat{x}_\um\sin\hat{x}_\um)}{\hat{x}_\um^2}-\tilde{\mu}_2\fNL\frac{6+6(\hat{x}_\um^2-1)\cos 2\hat{x}_\um-9\sin 2\hat{x}_\um}{5\hat{x}_\um^3},
}
which depends only on the combination $\tilde{\mu}_2\fNL$. Therefore, the maximal radius $\hat{x}_\um$ can be found as a function of $\tilde{\mu}_2\fNL$. We show its numerical solution in the top-left panel of Fig.~\ref{fig: xm and M fNL}.
Though we plot for small $\tilde{\mu}_2\fNL$ values, we note that the peak value $\hat{\zeta}(r=0)=\tilde{\mu}_2+\frac{3}{5}\fNL\tilde{\mu}_2^2$ is not increasing in the Gaussian kernel's amplitude $\tilde{\mu}_2$ for negatively large non-Gaussianity as $(3/5)\tilde{\mu}_2\fNL<-1/2$. The assumption that the peak position of $\zeta$ coincides with that of $\zeta^\uG$ is hence doubtful in that case.

Once the maximal radius is obtained, the mean compaction~\eqref{eq: barCm} can be calculated and Fig.~\ref{fig: mu vs fNL} shows its resultant contours.
By the blue line, we also show $\tilde{\mu}_{2,\uII}$ such that the perturbation for $\tilde{\mu}_2>\tilde{\mu}_{2,\uII}$ is type II, i.e., $\hat{R}^\prime(r)$ has a zero point.
For simplicity, we only consider the type I perturbation with $\tilde{\mu}_{2,\uth}<\tilde{\mu}_2<\tilde{\mu}_{2,\uII}$ where $\tilde{\mu}_{2,\uth}$ is the minimal $\tilde{\mu}_2$ such that the mean compaction $\hat{\bar{\calC}}_\um$ exceeds the threshold $\bar{\calC}_\uth=2/5$. It is shown by the red region in Fig.~\ref{fig: mu vs fNL}.
One sees that $\hat{\bar{\calC}}_\um$ can be smaller than $\bar{\calC}_\uth$ even for $\tilde{\mu}_2>\tilde{\mu}_{2,\uth}$ as we mentioned in the previous section in the negative $\fNL$ case. Particularly if $\fNL=\fNLmin\simeq-0.336$, $\hat{\bar{\calC}}_\um$ reaches the threshold $\bar{\calC}_\uth=2/5$ only at its maximal point $\tilde{\mu}_{2,\uth}(\fNLmin)$. However, the maximum $\hat{\calC}_\um$ itself is still increasing as shown in the top-right panel of Fig.~\ref{fig: xm and M fNL} and therefore it can be assumed to form a PBH with $\tilde{\mu}_2\geq\tilde{\mu}_{2,\uth}(\fNLmin)$.
For $\fNL<\fNLmin$, our treatment cannot find the type I PBH formation.
It may change in a different formulation but we leave the investigation of this region for future works and consider only $\fNL>\fNLmin$ in this work.
Note that $(3/5)\tilde{\mu}_2\fNL$ is smaller than $-1/2$ in the gray region. Therefore, the ``typical" curvature perturbation $\hat{\zeta}(r)$~\eqref{eq: typical zeta fNL} simply given by the Gaussian typical profile is not maximized at $r=0$ and hence it may fail to approximate the true typical non-Gaussian profile.

\bfe{width=0.7\hsize}{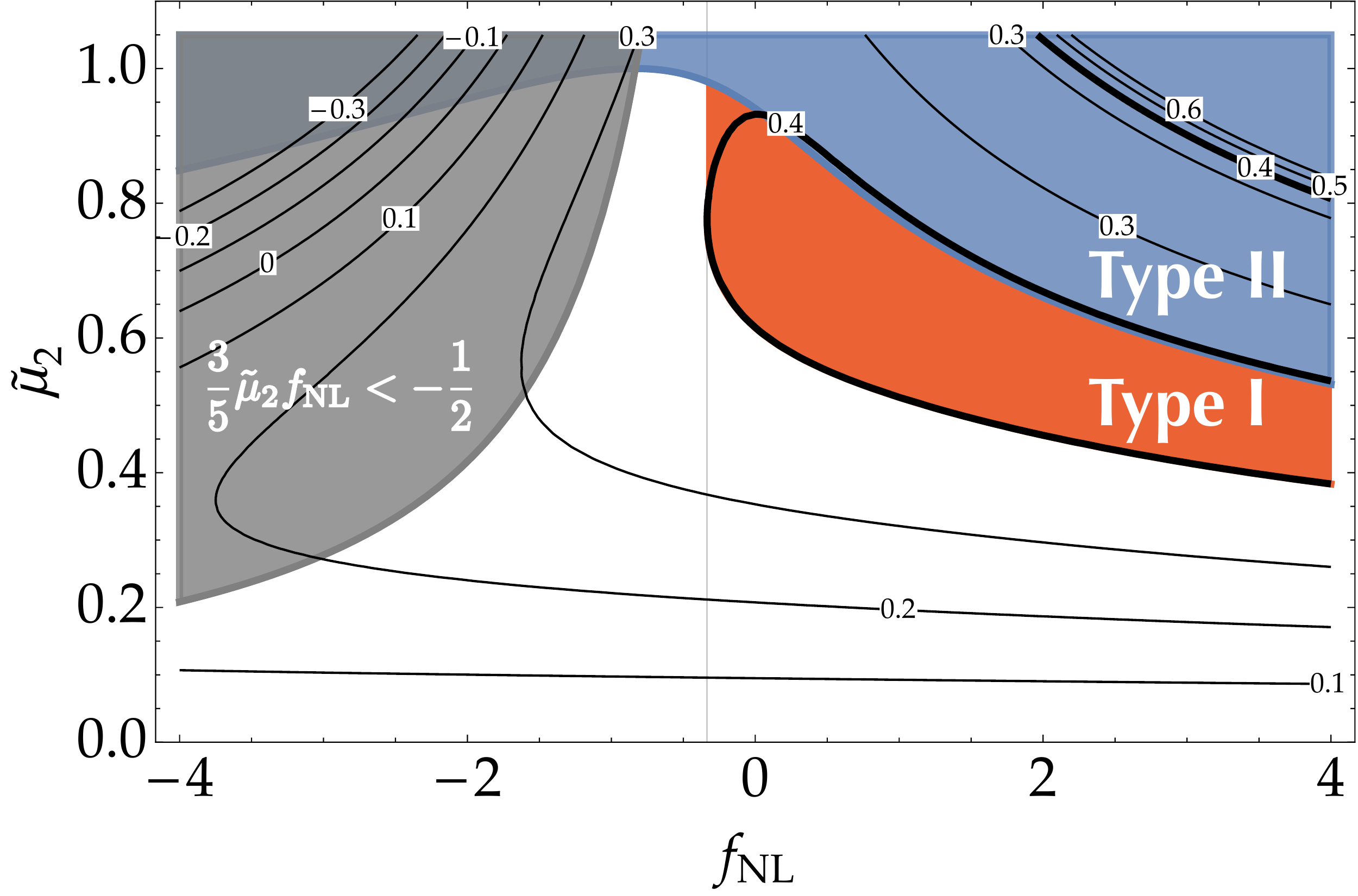}{Contours of the mean compaction $\hat{\bar{\calC}}_\um$. Thick lines correspond to the threshold value $\bar{\calC}_\uth=2/5$. In the blue region, the perturbation is type II, i.e., the areal radius $\hat{R}(r)$ is not monotonic. With the parameters in the red region, the type I PBH will be formed. The vertical line denotes $\fNLmin\simeq-0.336$ for $\fNL$ smaller than which no type I PBH is found in our treatment. In the gray region, the assumption that the typical non-Gaussian profile is simply given as a function of the typical Gaussian profile may be doubted.}{fig: mu vs fNL}

\begin{figure}
    \centering
    \begin{tabular}{c}
        \begin{minipage}{0.5\hsize}
            \centering
            \includegraphics[width=0.95\hsize]{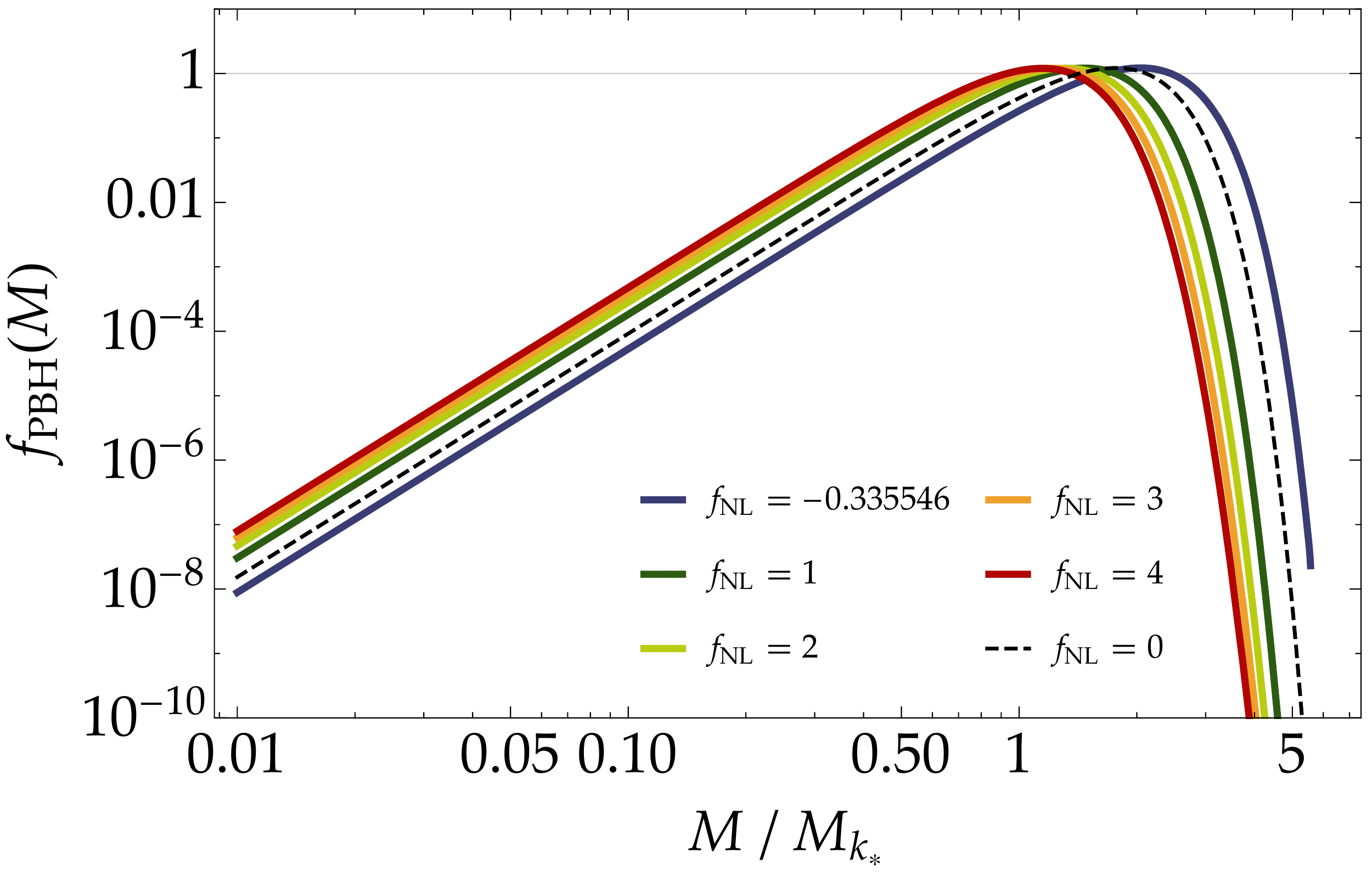}
        \end{minipage}
        \begin{minipage}{0.5\hsize}
            \centering
            \includegraphics[width=0.95\hsize]{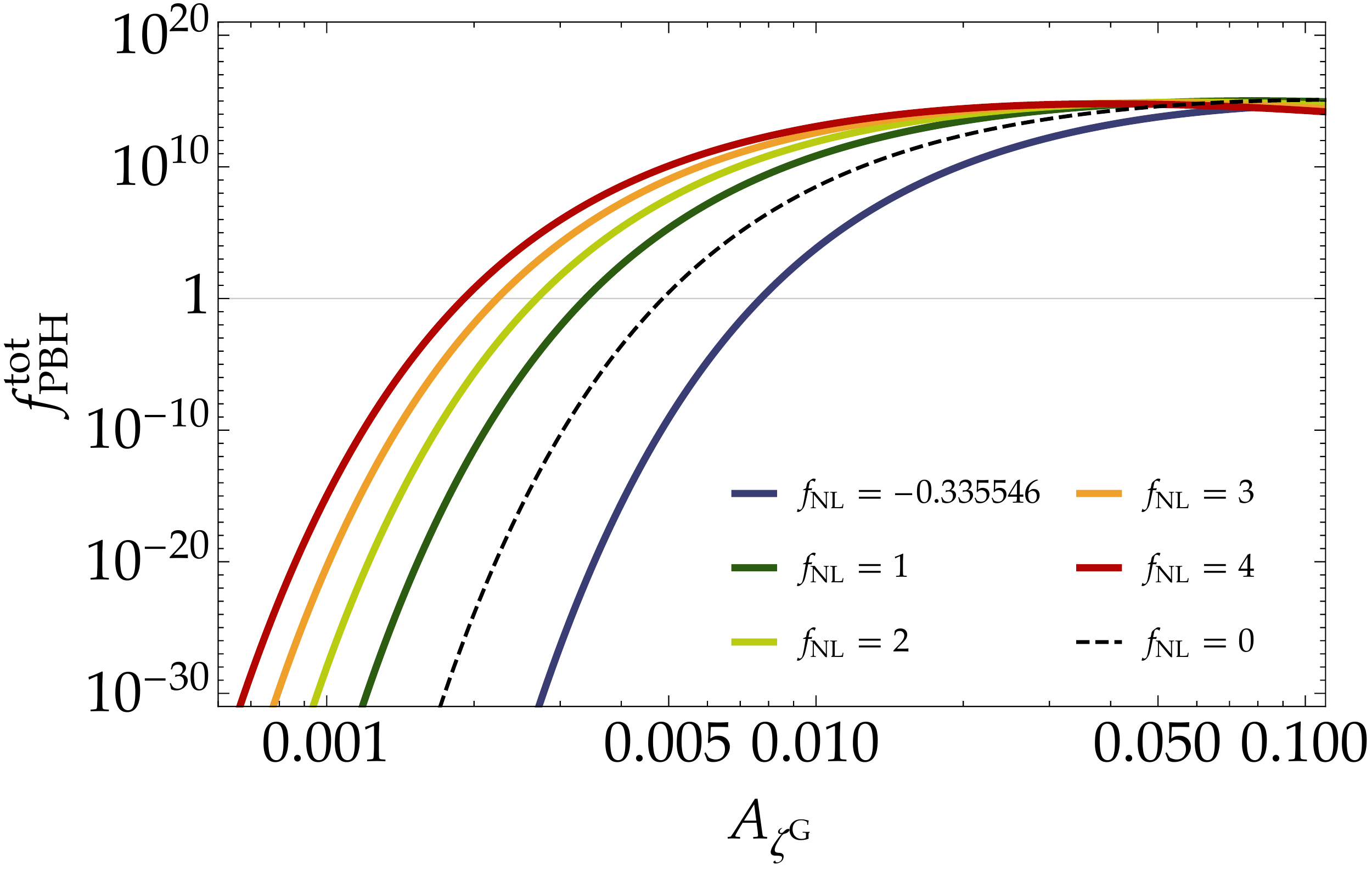}
        \end{minipage}
    \end{tabular}
    \caption{\emph{Left}: PBH mass functions for $k_*=1.56\times10^{13}\,\mathrm{Mpc}^{-1}$ with $(\fNL,\,A_{\zeta^\uG})=(\fNLmin\simeq-0.336,\,7.715\times10^{-3})$, $(1,\,3.381\times10^{-3})$, $(2,\,2.686\times10^{-3})$, $(3,\,2.230\times10^{-3})$, and $(4,\,1.908\times10^{-3})$ from right to left in comparison to the Gaussian case (dashed line) with $A_{\zeta^\uG}=4.863\times10^{-3}$. All mass functions are normalized so that $f_\PBH^\tot=1$. \emph{Right}: total PBH abundances $f_\PBH^\tot$ as functions of the amplitude $A_{\zeta^\uG}$ for various $\fNL$ values with the same color code to the left panel.}
    \label{fig: fPBH fNL}
\end{figure}

The current PBH abundance is given by the same equation~\eqref{eq: fNL Gauss} to the Gaussian case only with the difference in the mass-$\tilde{\mu}_2$ relation as
\bae{\label{eq: PBH mass fNL}
    M(\tilde{\mu}_2,\fNL)=\hat{x}_\um^2(\tilde{\mu}_2\fNL)\ee^{2\hat{\zeta}_\um(\tilde{\mu}_2,\fNL)}K\pqty{\tilde{\mu}_2-\tilde{\mu}_{2,\uth}(\fNL)}^\gamma M_k(k_*),
}
which is plotted in the bottom panel of Fig.~\ref{fig: xm and M fNL}.
The resultant mass spectra with the tuned $A_{\zeta^\uG}$ such that $f_\PBH^\tot=1$ and the total PBH abundance $f_\PBH^\tot$ as a function of $A_{\zeta^\uG}$ are shown in Fig.~\ref{fig: fPBH fNL}.
The right panel indicates that larger $\fNL$ helps the PBH formation with a smaller amplitude $A_{\zeta^\uG}$ as expected. Also, the mass spectrum is slightly varied even if the total abundance $f_\PBH^\tot$ is fixed to unity.

\subsection{Exponential tail}\label{sec: exponential tail}

Recently, another local-type non-Gaussianity in the ultra slow-roll inflation model is attracting attentions.
In the ultra slow-roll limit (i.e., the exactly flat inflaton potential and the almost constant Hubble parameter), the $\delta N$ formalism reveals that the curvature perturbation $\zeta(\bfx)$ is related to the Gaussian field $\zeta^\uG(\bfx)$ by (see, e.g., Refs.~\cite{Cai:2017bxr,Atal:2019cdz,Atal:2019erb,Biagetti:2021eep})
\bae{\label{eq: zeta exp tail}
    \zeta(\bfx)=-\frac{1}{3}\ln(1-3\zeta^\uG(\bfx)),
}
neglecting the resummation effect of the inflaton perturbations. It is not well-defined for $\zeta^\uG\geq1/3$ and the resummation effect should be correctly involved with use of, e.g., the stochastic formalism in such a case, though we simply neglect perturbations larger than $1/3$ in this paper.\footnote{In our exact USR case, $\fNL=5/2$, Ref.~\cite{Atal:2019erb} discusses that the PBH formation corresponds to $\zeta^\uG\geq1/3$ is in fact rarer than that with $\zeta^\uG<1/3$.}
Note that this expression is consistent with the well-known result $\fNL=5/2$ (see, e.g., Ref.~\cite{Namjoo:2012aa}) in the ultra slow-roll model up to the quadratic order as
\bae{
    \zeta(\bfx)=\zeta^\uG(\bfx)+\frac{3}{5}\times\frac{5}{2}\pqty{\zeta^\uG(\bfx)}^2+\calO\pqty{\pqty{\zeta^\uG(\bfx)}^3}.
}
The important feature of Eq.~\eqref{eq: zeta exp tail} can be seen in the probability density function (PDF) of $\zeta$. The PDF is given by
\bae{
    P(\zeta)=\abs{\dv{\zeta^\uG}{\zeta}}P(\zeta^\uG)=\frac{1}{\sqrt{2\pi\sigma_0^2}}\exp\bqty{-3\zeta-\frac{(1-\ee^{-3\zeta})^2}{18\sigma_0^2}},
}
through the Gaussian distribution 
\bae{
    P(\zeta^\uG)=\frac{1}{2\pi\sigma_0^2}\ee^{-{\zeta^\uG}^2/(2\sigma_0^2)}.
}
In the large $\zeta$ limit, it only decays in an exponential way $P(\zeta)\propto\ee^{-3\zeta}$ instead of the Gaussian suppression. 
Such an exponential tail is also confirmed in the stochastic approach~\cite{Pattison:2017mbe,Ezquiaga:2019ftu,Figueroa:2020jkf,Pattison:2021oen}.
The PBH abundance is therefore expected to be much amplified compared to the simple $\fNL$ correction with $\fNL=5/2$.
In this subsection, we consider the PBH formation via this exponential-tail type non-Gaussianity~\eqref{eq: zeta exp tail}.

\begin{figure}
    \centering
    \begin{tabular}{c}
        \begin{minipage}{0.513\hsize}
            \centering
            \includegraphics[width=0.95\hsize]{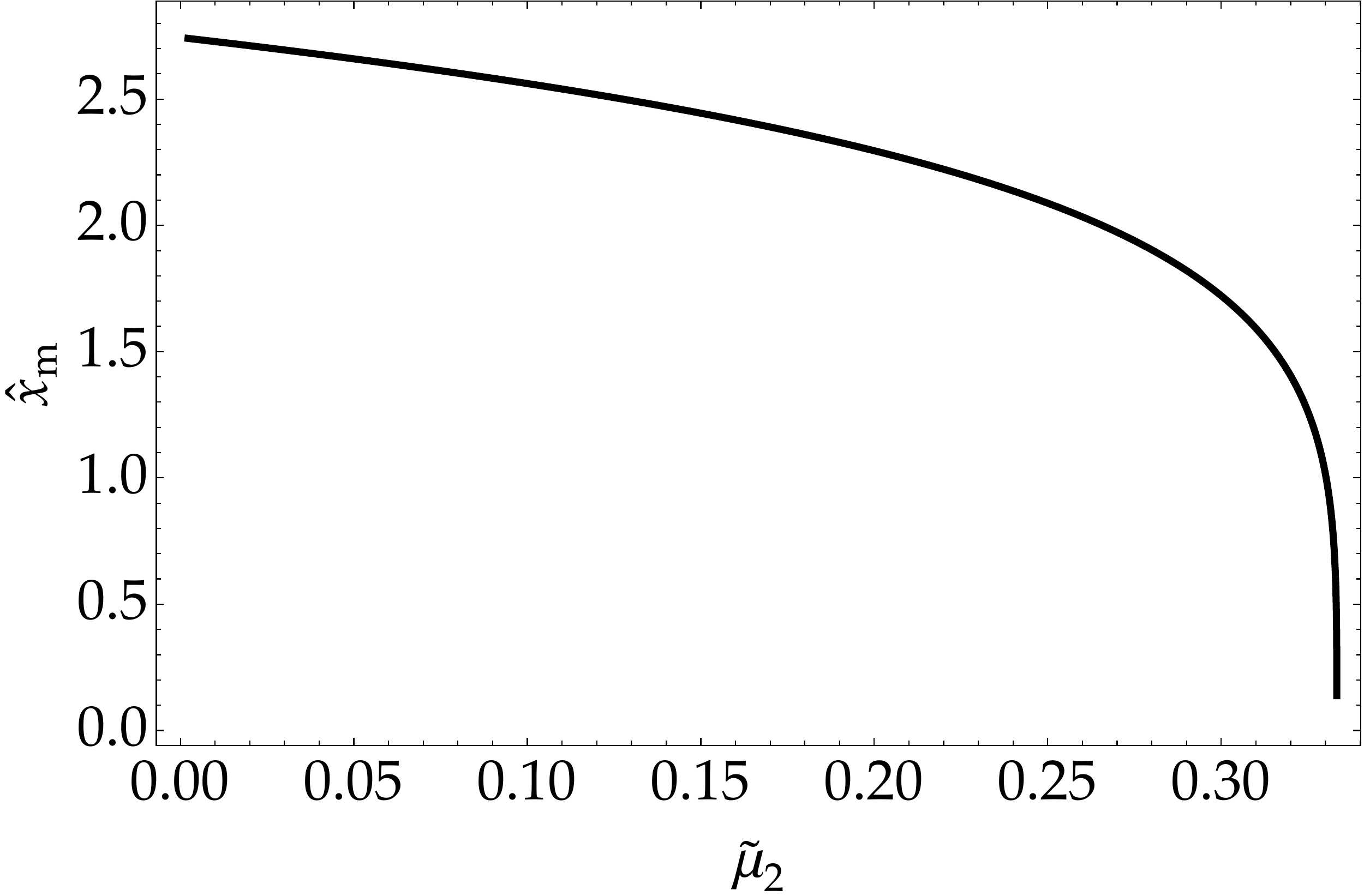}
        \end{minipage}
        \begin{minipage}{0.487\hsize}
            \centering
            \includegraphics[width=0.95\hsize]{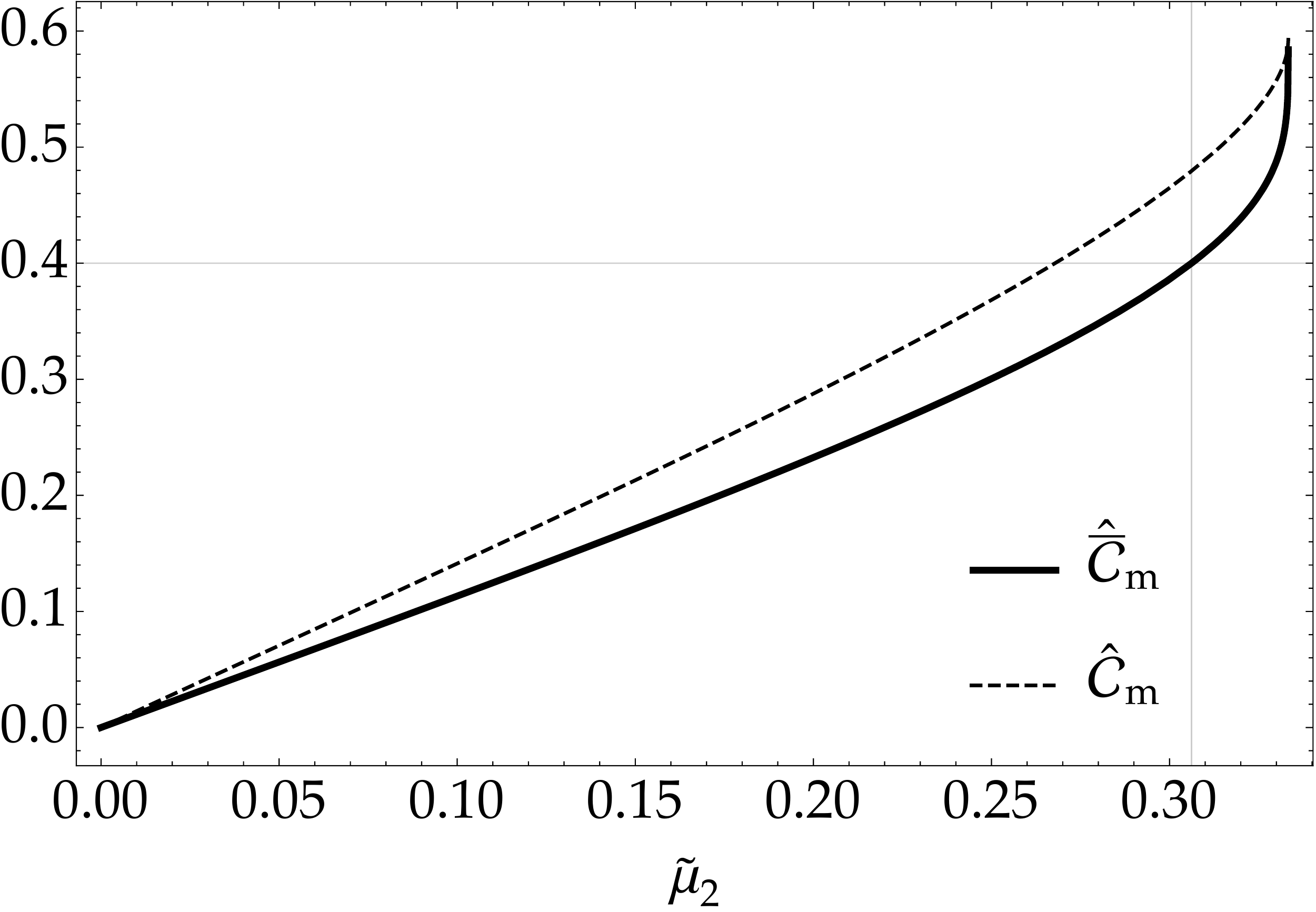}
        \end{minipage} \\ 
        \begin{minipage}{0.5\hsize}
            \centering
            \includegraphics[width=0.95\hsize]{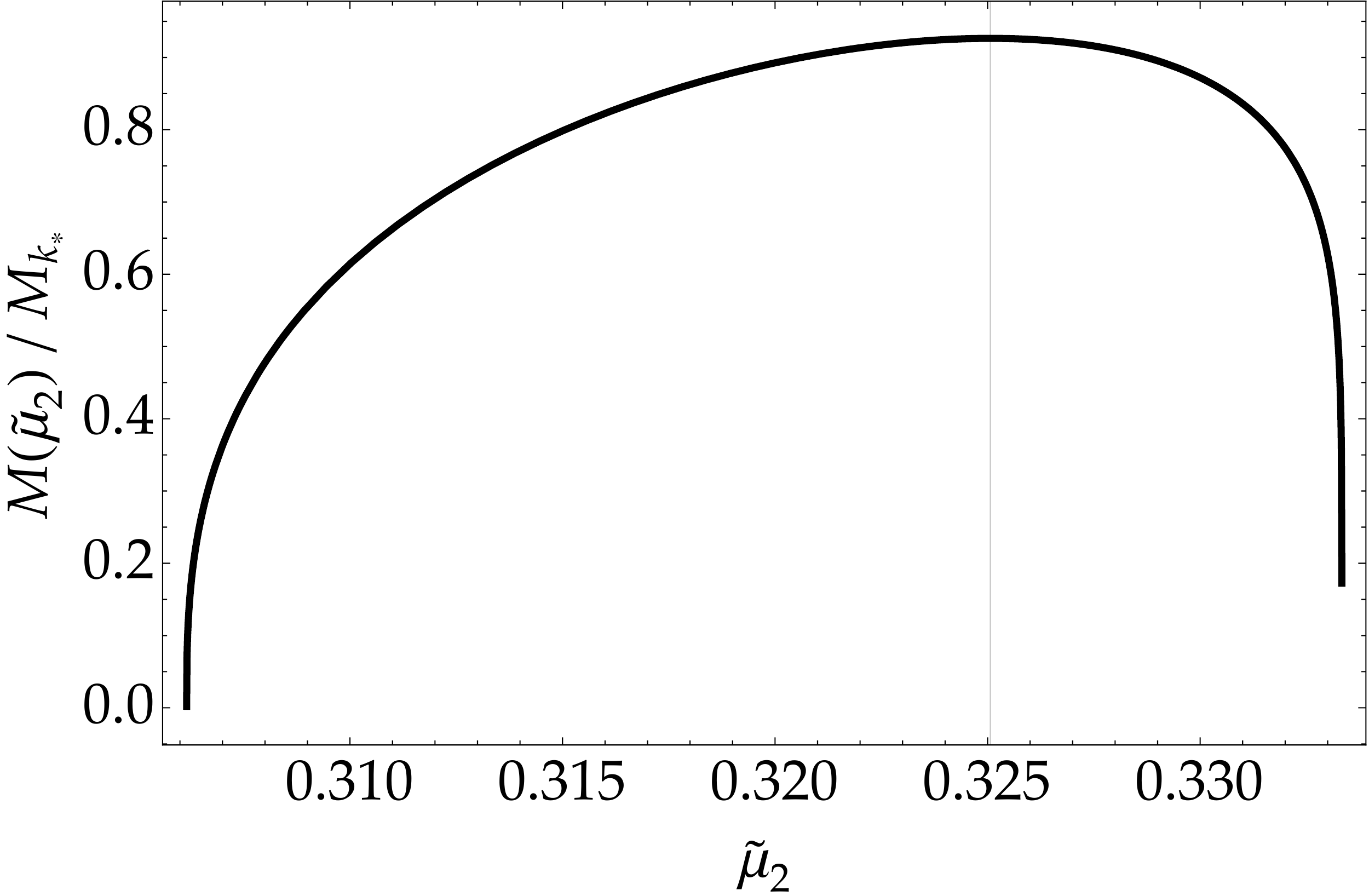}
        \end{minipage}
    \end{tabular}
    \caption{\emph{Top-Left}: the maximal radius $\hat{x}_\um$. \emph{Top-Right}: 
    the mean compaction $\hat{\bar{\calC}}_\um$ (thick) and the compaction maximum $\hat{\calC}_\um$ (dashed). The horizontal thin line is $\bar{\calC}_\uth=2/5$ and the vertical one is the corresponding threshold $\tilde{\mu}_{2,\uth}\simeq0.306$. \emph{Bottom}: the resultant PBH mass~\eqref{eq: PBH mass exp} with respect to $\tilde{\mu}_2$. It is maximized at $\tilde{\mu}_{2,\max}\simeq0.325$ indicated by the vertical thin line.}
    \label{fig: x and C exp}
\end{figure}

The procedure is similar to the $\fNL$ expansion case. One first finds that the perturbation is type I for $\tilde{\mu}_2<1/3$.
Then the maximal radius $\hat{x}_\um$ and the mean compaction $\hat{\bar{\calC}}_\um$ can be obtained as shown in the top panels of Fig.~\ref{fig: x and C exp}.
The threshold is translated to the $\tilde{\mu}_2$'s language as $\tilde{\mu}_{2,\uth}\simeq0.306$.
The PBH mass
\bae{\label{eq: PBH mass exp}
    M(\tilde{\mu}_2)=\hat{x}_\um^2(\tilde{\mu}_2)\ee^{2\hat{\zeta}_\um(\tilde{\mu}_2)}K(\tilde{\mu}_2-\tilde{\mu}_{2,\uth})^\gamma M_k(k_*),
}
is exhibited in the bottom panel of Fig.~\ref{fig: x and C exp}. Interestingly, in the exponential tail case, the PBH mass is not monotonically increasing with respect to the perturbation amplitude $\tilde{\mu}_2$ but takes its maximum at $\tilde{\mu}_2=\tilde{\mu}_{2,\max}\simeq0.325$.
That is, a too large perturbation corresponds to small PBH mass, and it is caused by the strong shrink of the maximal radius $\hat{x}_\um$ as can be seen in the top-left panel of Fig.~\ref{fig: x and C exp}. Due to the divergent behavior of $\zeta$ toward $\zeta^\uG\to1/3$, large perturbations have cuspy profiles around the origin $r=0$ and cause the shrink of $\hat{x}_\um$.
This non-monotonicity of $M(\tilde{\mu}_2)$ means that its inverse $\tilde{\mu}_2(M)$ is two-valued. Therefore the PBH abundance with the mass $M$ can be contributed both by the smaller $\tilde{\mu}_2(M)$ and the larger $\tilde{\mu}_2$.
Practically, the contributions from the larger $\tilde{\mu}_2$ are however suppressed probabilistically, so that we simply neglect them and restrict $\tilde{\mu}_2$ to the range of $\tilde{\mu}_{2,\uth}<\tilde{\mu}_2<\tilde{\mu}_{2,\max}$.

\begin{figure}
    \centering
    \begin{tabular}{c}
        \begin{minipage}{0.5\hsize}
            \centering
            \includegraphics[width=0.95\hsize]{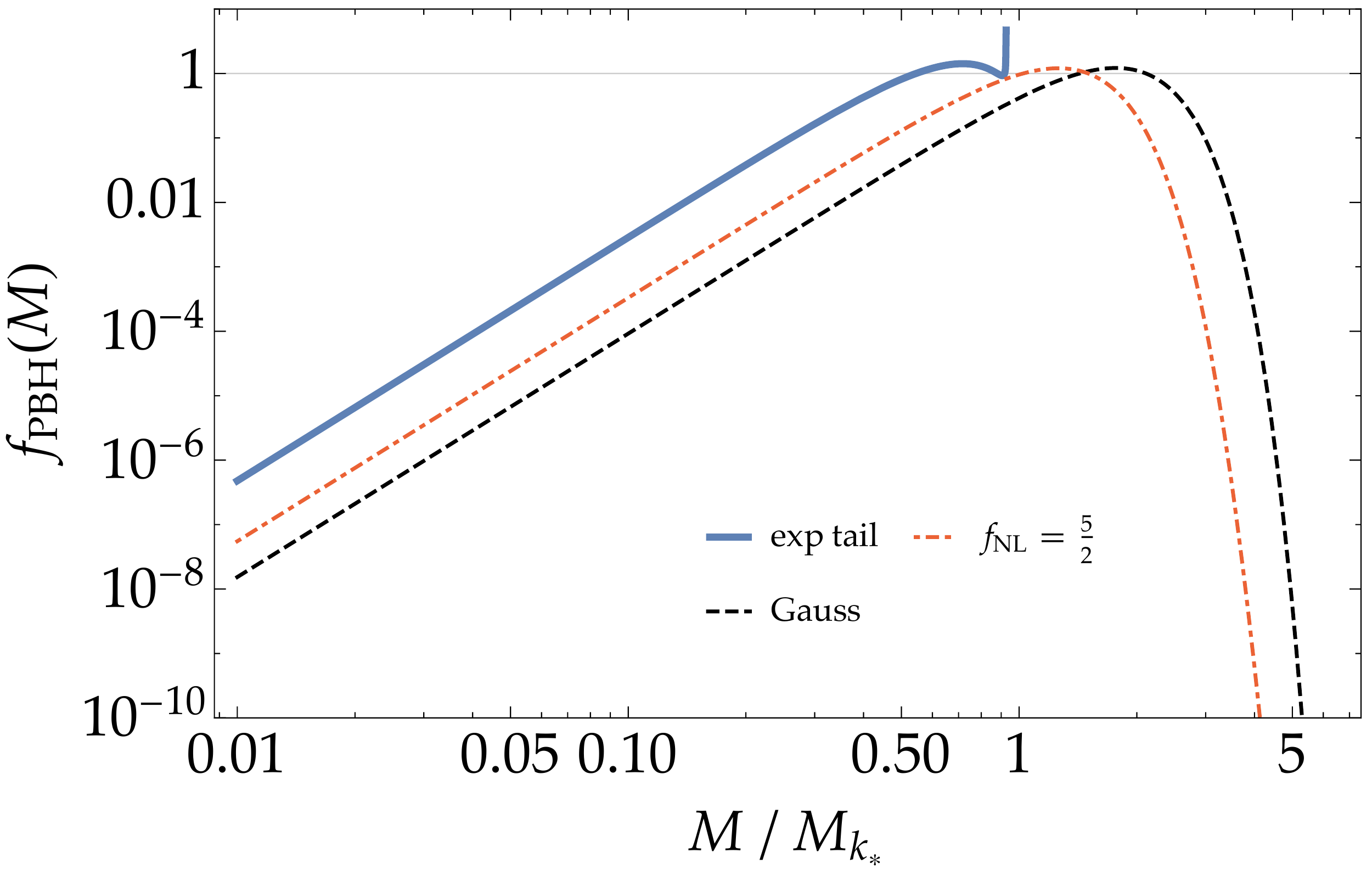}
        \end{minipage}
        \begin{minipage}{0.5\hsize}
            \centering
            \includegraphics[width=0.95\hsize]{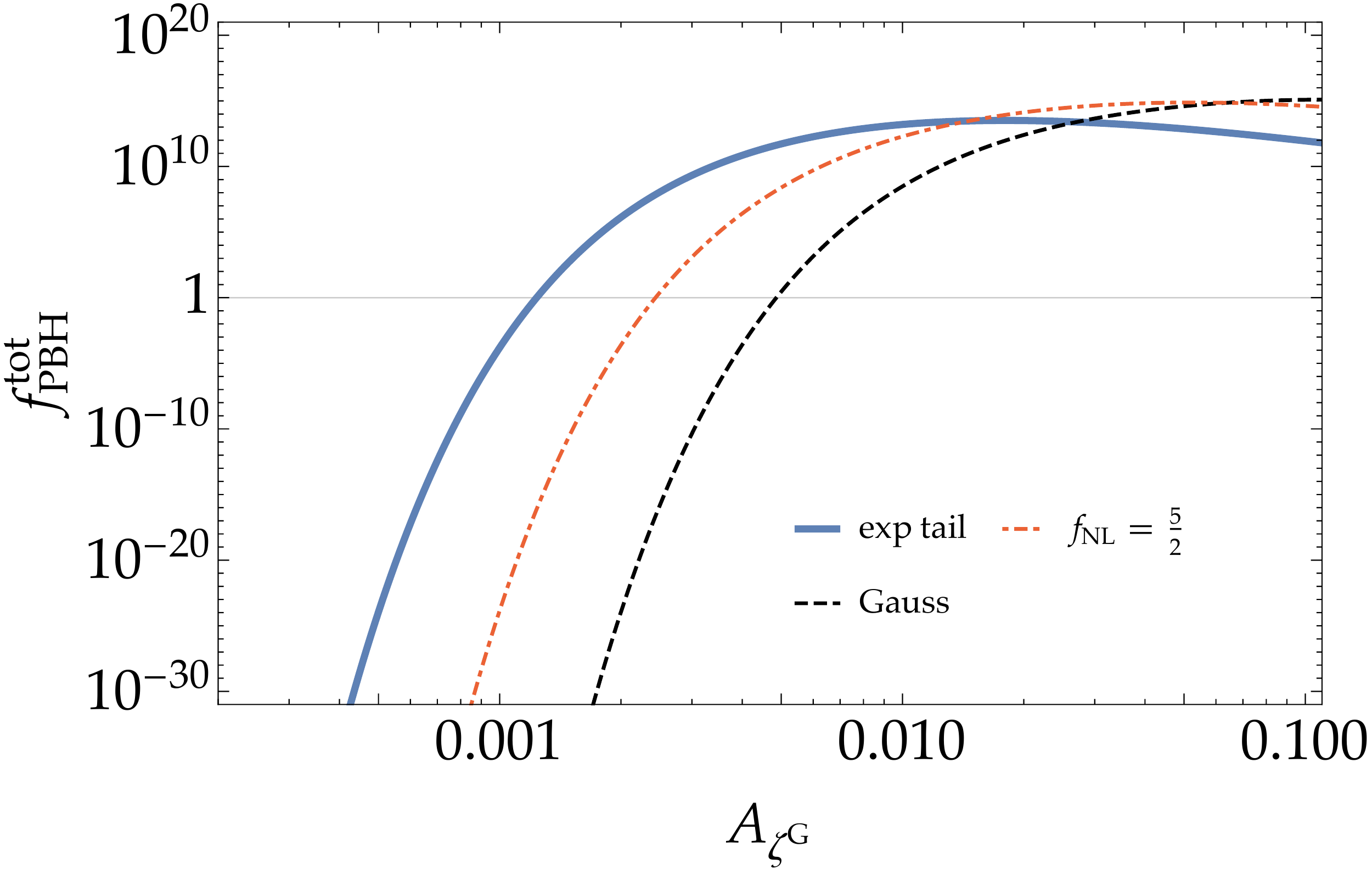}
        \end{minipage}
    \end{tabular}
    \caption{\emph{Left}: PBH mass functions with $A_{\zeta^\uG}=1.235\times10^{-3}$, $2.437\times10^{-3}$, and $4.863\times10^{-3}$ for the exponential tail (blue), $\fNL=5/2$ (red dot-dashed) and the Gaussian (black dashed) curvature perturbation, respectively, with $k_*=1.56\times10^{13}\,\mathrm{Mpc}^{-1}$. \emph{Right}: total PBH abundances $f^\tot_\PBH$ as functions of $A_{\zeta^\uG}$ with the same plot style to the left panel.}
    \label{fig: fPBH exp}
\end{figure}

The resultant PBH mass spectra are shown in Fig.~\ref{fig: fPBH exp} in comparison to the Gaussian case as well as the simple $\fNL$ expansion with $\fNL=5/2$.
As expected, the right panel indicates that the exponential tail increases the PBH abundance compared even to the $\fNL=5/2$ case. Also, the mass function is characteristic with a hard cut at the maximum mass $M(\tilde{\mu}_{2,\max})$. Though it shows an apparent divergence there, the integral $f_\PBH^\tot$ is regularized.

\section{Discussion and Conclusions}\label{sec: discussion and conclusions}

In this paper, we update the peak theory for the PBH abundance based on Refs.~\cite{Yoo:2018kvb,Yoo:2019pma,Yoo:2020dkz} and apply it to the exponential tail non-Gaussianity, which is characteristic in ultra slow-roll models.
The features of our peak theory are summarized as follows:
\begin{itemize}
    \item It treats the peak of the Laplacian of the curvature perturbation, $-\Delta\zeta$, to avoid the long-wavelength contamination, following Ref.~\cite{Yoo:2020dkz}.
    \item The local-type non-Gaussianity including the exponential tail is taken into account with the assumption that the peak position of the non-Gaussian field corresponds to that of the Gaussian kernel field, following Ref.~\cite{Yoo:2019pma}.
    \item We adopt the scaling relation for the PBH mass~\eqref{eq: MPBH}.
    \item For the threshold judgment, we employ the averaged compaction function~\eqref{eq: barCm} to relax the profile dependence~\cite{Escriva:2019phb,Atal:2019erb}.
\end{itemize}
The PBH mass function and abundance are exemplified in Fig.~\ref{fig: fPBH Gauss} for Gaussian, in Fig.~\ref{fig: fPBH fNL} for $\fNL$ non-Gaussian, and in Fig.~\ref{fig: fPBH exp} for exponential tail curvature perturbations with a monochromatic power spectrum~\eqref{eq: monochromatic power}.
In the exponential tail case, we found that the PBH abundance is much enhanced, even compared to the corresponding $\fNL=5/2$ expansion, as expected, and also the mass function has a characteristic maximal mass, i.e., a hard cutoff rather than the exponential suppression, which is a consequence of a cuspy peak profile of the curvature perturbation. It is not seen in the simple Press--Schechter approach.

The formulation of the peak theory given in Sec.~\ref{sec: peak theory} can be applied to another narrow-peak power spectrum than the exact monochromatic one without difficulty.
In the case where the power spectrum is broad, one needs a UV cutoff $k_\uW$ not only to regularize the variance $\sigma_n$ but also to properly evaluate the abundance of large PBHs.\footnote{Even an IR cut may be also required to $\sigma_0$ for an almost scale-invariant power spectrum. In a realistic scenario, it will be automatically introduced anyway because the power spectrum should be reduced to $\sim10^{-9}$ on a smaller scale than the CMB's one $\sim1\,\mathrm{Mpc}$ at largest.}
This is because the peak profile can be contaminated by short-wavelength fluctuations as the PBH formation is judged by the innermost extremal radius (note that the long-wavelength contamination is removed by treating the Laplacian of the curvature perturbation instead of the curvature perturbation itself), and the large mass PBH abundance would be underestimated in our formalism.
Hence a window function $W(k/k_\uW)$ should be introduced in the variance and two point function as
\beae{
    &\sigma_n^2(k_\uW)=\int\frac{\dd{k}}{k}W\pqty{\frac{k}{k_\uW}}k^{2n}\calP_{\zeta^\uG}(k), \\
    &\psi_n(r;k_\uW)=\frac{1}{\sigma_n^2(k_\uW)}\int\frac{\dd{k}}{k}W\pqty{\frac{k}{k_\uW}}k^{2n}\frac{\sin(kr)}{kr}\calP_{\zeta^\uG}(k),
}
to omit the short-wavelength modes $k\gg k_\uW$. This is expected also in a real universe after the horizon reentry of the short-wavelength modes because the subhorizon perturbations in the radiation-dominated era are soon smoothed away and do not contribute to the compaction function.
Among various types of window functions, Ref.~\cite{Yoo:2020dkz} recommends the Fourier-space top-hat one $W^\mathrm{FTH}(k/k_\uW)=\theta(k_\uW-k)$ to minimize an artificial reduction of perturbations of interest ($k<k_\uW$) due to the window function.
Once $k_\uW$ is fixed, the same procedure leads to the PBH abundance $f_\PBH(M;k_\uW)$ under a certain window condition, and the true PBH abundance can be understood as its maximum when $k_\uW$ is varied in a relevant range: e.g.,
\bae{
    f_\PBH(M)=\max\Bqty{f_\PBH(M;k_\uW)\mid k_\uW\in\mathbb{R}^+}.
}

We also note that the exponential tail can also be realized as the $\chi^2$-distribution (if $\chi$ is Gaussian, $\chi^2$ decays only exponentially), other than the ultra slow-roll model.
It is the case such that the curvature perturbation is sourced by the second order effect of some Gaussian field.
Indeed an exponential tail behavior has been seen for the curvature perturbation from $(\text{tensor})\times(\text{tensor})$~\cite{Nakama:2016enz} and from $(\text{vector})\times(\text{vector})$~\cite{Saga:2020ics} (see also Refs.~\cite{Garcia-Bellido:2016dkw,Garcia-Bellido:2017aan}).
However, unlike the ultra slow-roll case, the shape of the probability distribution function around the peak also deviates from that of the Gaussian distribution, largely.
Thus, for this issue, further improvements in the formulation may be needed.
Anyway, these cases may also be interesting examples of the peak theory with a non-Gaussian tail, and we leave them for future works.

\acknowledgments

This work is supported by JSPS KAKENHI Grant Numbers
JP19H01894 (N.K.), JP20H01894 (N.K.), JP20H05851 (N.K.), JP21H01078 (N.K.), 
JP19K14707 (Y.T.), JP21K13918 (Y.T.),
JP19H01895 (C.Y.), JP20H05850 (C.Y.), and JP20H05853 (C.Y.).
S.Y. is supported by JSPS Grant-in-Aid for Scientific Research 
(B) No. JP20H01932 and (C) No. JP20K03968.

\appendix

\section{Comparison to the (latest) Press--Schechter approach}\label{sec: comparison to other approaches}

The concept of the Press--Schechter approach is to estimate the formation rate of the collapsed object by the probability that some coarse-grained random field exceeds the threshold value.
For PBH, it seems the current trend~\cite{Kawasaki:2019mbl,Biagetti:2021eep} to use the compaction function itself (not averaged one used in our peak theory) as this random field because i) it is related to the curvature perturbation by the simple analytic equation~\eqref{eq: compaction C} and ii) it is already understood as the coarse-grained density contrast with the real-space top-hat window as discussed in Eq.~\eqref{eq: calC and delta}.
Fig.~\ref{fig: Cm and M Gauss} shows that $\calC_\uth=\calC(\tilde{\mu}_{2,\uth})\simeq0.587$ corresponds to our threshold $\tilde{\mu}_{2,\uth}\simeq0.615$ in the monochromatic and Gaussian case, and we adopt this value as a universal threshold.
Noting that the compaction function~\eqref{eq: compaction C} is a function only of $\delta_l=-\frac{4}{3}r\zeta^\prime(r)$ as
\bae{
    \calC=\delta_l-\frac{3}{8}\delta_l^2 \qc
    \delta_l=-\frac{4}{3}r\zeta^\prime(r),
}
one can convert the compaction's threshold $\calC_\uth\simeq0.587$ to that in $\delta_l$ as
\bae{
    \delta_{l,\uth}=\frac{4}{3}\pqty{1-\sqrt{1-\frac{3}{2}\calC_\uth}}\simeq0.872.
}
The spherical symmetry of $\zeta$ has been assumed.
Obviously, the compaction function gets decreasing for $\delta_l>\delta_{l,\max}=4/3$, corresponding to the type II perturbations. Hereafter we restrict ourselves to $\delta_l\leq\delta_{l,\max}$ for simplicity.

Let us then consider the PDF of $\delta_l$.
In the case of $\fNL$ expansion~\eqref{eq: fNL expansion} first, $\delta_l$ is expressed as
\bae{
    \delta_l=XY \qc X=-\frac{4}{3}r{\zeta^\uG}^\prime(r) \qc Y=1+\frac{6}{5}\fNL\zeta^\uG.
}
$X$ and $Y$ are Gaussian variables whose covariance matrix $\Sigma=\spmqty{\sigma_X^2 & \sigma_{XY}^2 \\ \sigma_{YX}^2 & \sigma_Y^2}$ is computed as\footnote{One may insert the transfer function of the radiation-dominated universe $T^2(k\eta)=W^2(k\eta/\sqrt{3})$ at the horizon cross of $R(r)$, i.e. $\eta=R(r)$, into these integrals to cut the subhorizon modes. Here we do not care about it because we only consider the monochromatic case: $\calP_{\zeta^\uG}(k)=A_{\zeta^\uG}\delta(\ln k-\ln k_*)$.}
\beae{
    &\sigma_X^2=\braket{X^2}=\pqty{\frac{4}{9}}^2\int(kr)^4W^2(kr)\calP_{\zeta^\uG}(k)\dd{\ln k}, \\
    &\sigma_{XY}^2=\sigma_{YX}^2=\braket{X(Y-1)}=\frac{4}{9}\times\frac{6}{5}\fNL\int(kr)^2W(kr)\frac{\sin kr}{kr}\calP_{\zeta^\uG}(k)\dd{\ln k}, \\
    &\sigma_Y^2=\braket{(Y-1)^2}=\pqty{\frac{6}{5}\fNL}^2\int\frac{\sin^2kr}{(kr)^2}\calP_{\zeta^\uG}(k)\dd{\ln k},
}
with the Fourier transform of the real-space top-hat window $W(z)=\frac{3}{z^3}(\sin z-z\cos z)$, making use of the spherical expansion~\cite{Kawasaki:2019mbl}
\bae{
    \zeta(r)=\frac{1}{\sqrt{2}\pi}\int\dd{k}k\frac{\sin kr}{kr}\zeta_k \qc 
    \braket{\zeta_k\zeta_{k^\prime}}=\delta(k-k^\prime)\frac{2\pi^2}{k^3}\calP_{\zeta^\uG}(k).
}
The joint probability of $X$ and $Y$ are then given by
\bae{
    P(X,Y)=\frac{1}{2\pi\sqrt{\det \Sigma}}\exp\bqty{-\frac{1}{2}\pmqty{X, & Y-1}\Sigma^{-1}\pmqty{X \\ Y-1}},
}
and the PDF of $\delta_l$ can be computed as
\bae{
    P^{\fNL}(\delta_l)=\int\dd{Y}P\pqty{X=\frac{\delta_l}{Y},Y}.
}
On the other exponential-tail side~\eqref{eq: zeta exp tail}, $\delta_l$ is given by
\bae{
    \delta_l=\frac{X}{Y^\prime} \qc Y^\prime=1-3\zeta^\uG.
}
The covariance matrix $\Sigma^\prime=\spmqty{\sigma_X^2 & \sigma_{XY^\prime}^2 \\ \sigma_{Y^\prime X} & \sigma_{Y^\prime}^2}$ reads
\beae{
    &\sigma_{XY^\prime}^2=\sigma_{Y^\prime X}^2=\braket{X(1-Y^\prime)}=\frac{4}{9}\times3\int(kr)^2W(kr)\frac{\sin kr}{kr}\calP_{\zeta^\uG}(k)\dd{\ln k}, \\
    &\sigma_{Y^\prime}^2=\braket{(1-Y^\prime)^2}=3^2\int\frac{\sin^2kr}{(kr)^2}\calP_{\zeta^\uG}(k)\dd{\ln k}.
}
The joint probability of $X$ and $Y^\prime$ and the PDF of $\delta_l$ are expressed as
\beae{
    &P(X,Y^\prime)=\frac{1}{2\pi\sqrt{\det \Sigma^\prime}}\exp\bqty{-\frac{1}{2}\pmqty{X, & 1-Y^\prime}{\Sigma^\prime}^{-1}\pmqty{X \\ 1-Y^\prime}}, \\
    &P^\mathrm{exp}(\delta_l)=\int\dd{Y^\prime}P\pqty{X=\delta_lY^\prime,Y^\prime}.
}
Particularly in the monochromatic case, $\calP_{\zeta^\uG}(k)=A_{\zeta^\uG}\delta(\ln k-\ln k_*)$, these expressions are much simplified because $Y$ and $Y^\prime$ are dependent on $X$ as one can obviously see $\sigma_{XY}^2=\sigma_X\sigma_Y$ and $\sigma_{XY^\prime}^2=\sigma_X\sigma_{Y^\prime}$ and thus $\det\Sigma=\det\Sigma^\prime=0$.
In this case, $Y$ and $Y^\prime$ are given by $Y=1+\frac{\sigma_Y}{\sigma_X}X$ and $Y^\prime=1-\frac{\sigma_{Y^\prime}}{\sigma_X}X$. Therefore the PDF of $\delta_l$ reads
\beae{
    &P^{\fNL}(\delta_l)=P_\uG\pqty{X=\frac{-\sigma_X\pm\sqrt{\sigma_X^2+4\sigma_X\sigma_Y\delta_l}}{2\sigma_Y},\sigma_X^2}, \\
    &P^\mathrm{exp}(\delta_l)=P_\uG\pqty{X=\frac{\sigma_X\delta_l}{\sigma_X+\sigma_{Y^\prime}\delta_l},\sigma_X^2},
}
with
\bae{
    \sigma_X=\frac{4}{9}(k_*r)^2W(k_*r)\sqrt{A_{\zeta^\uG}} \qc
    \sigma_Y=\frac{6}{5}\fNL\frac{\sin k_*r}{k_*r}\sqrt{A_{\zeta^\uG}} \qc
    \sigma_{Y^\prime}=3\frac{\sin k_*r}{k_*r}\sqrt{A_{\zeta^\uG}},
}
and the Gaussian distribution $P_\uG(x,\sigma^2)=\frac{1}{\sqrt{2\pi\sigma^2}}\ee^{-x^2/(2\sigma^2)}$.

In general, the mass function should be investigated by varying the coarse-graining scale $R(r)$. However, in the monochromatic case, it is often fixed to $k_*r=x_\um\simeq2.74$ for simplicity, which maximizes the compaction function in the monochromatic and Gaussian case.
Furthermore, the horizon cross condition $a\ee^{\zeta}rH=1$ is simplified to $arH=1$, neglecting the $\ee^\zeta$ factor.
Thus the PBH mass is approximated by
\bae{
    M(\delta_l)\sim K\bqty{\pqty{\delta_l-\frac{3}{8}\delta_l^2}-\calC_\uth}^\gamma M_k(k_*/x_\um)=x_\um^2K\bqty{\pqty{\delta_l-\frac{3}{8}\delta_l^2}-\calC_\uth}^\gamma M_{k_*},
}
expressing the critical behavior in terms of the compaction function.\footnote{While the power $\gamma\simeq0.36$ is universal, the coefficient $K$ depends not only on the peak profile but also on the choice of the scaling parameter, strictly speaking, as they (e.g., $K_\calC$ in $\calC$ and $K_{\tilde{\mu}_2}$ in $\tilde{\mu}_2$) are related by the expansion around the threshold $K_\calC\pqty{\hat{\calC}_\um^\prime(\tilde{\mu}_{2,\uth})}^\gamma=K_{\tilde{\mu}_2}$. $K_\calC\simeq1.3$ corresponds to our $K_{\tilde{\mu}_2}\simeq1$ in the monochromatic and Gaussian case, while $K_\calC\simeq3.3$ is often used in the literature~\cite{Choptuik:1992jv,Evans:1994pj,Niemeyer:1997mt}. In this paper, we uniformly approximate them by $K\simeq1$ for simplicity, neglecting these subtleties, as they are anyway uncertain, depending on the peak profile.}
Accordingly, at the horizon reentry $arH=1$, each Hubble patch is converted into a PBH of mass $[M(\delta_l),M(\delta_l)\ee^{\dd{\ln M}}]$ with the probability $P(\delta_l)\abs{\dv{\ln M}{\delta_l}}^{-1}$.
Therefore the energy ratio of PBHs to the background radiation at their formation time is given by
\bae{
    \beta(M)\dd{\ln M}&=\frac{\rho_\PBH(M)}{\bar{\rho}}\dd{\ln M}=\frac{M}{M_k(k_*/x_\um)}\abs{\dv{\ln M}{\delta_l}}^{-1}P(\delta_l)\dd{\ln M} \nonumber \\
    &\sim\frac{K\bqty{\pqty{\delta_l-\frac{3}{8}\delta_l^2}-\calC_\uth}^{\gamma+1}}{\gamma\pqty{1-\frac{3}{4}\delta_l}}P(\delta_l)\dd{\ln M}.
}
The current PBH abundance~\eqref{eq: fPBH} then reads (see, e.g., Ref.~\cite{Tada:2019amh})
\bae{
    \!\!\!\!\!\!
    f_\PBH(M)&\!=\frac{\Omega_\um h^2}{\Omega_\DM h^2}\frac{T(x_\um/k_*)}{T_\eq}\beta(M) \nonumber \\
    &\!\simeq\!\pqty{\frac{\Omega_\DM h^2}{0.12}}^{\!\!-1}\!\pqty{\frac{g_*}{106.75}}^{\!1/6}\!\pqty{\frac{x_\um}{2.74}}^{\!-1}\!\pqty{\frac{k_*}{1.56\times10^{13}\,\mathrm{Mpc}^{-1}}}\pqty{\frac{\beta(M)}{6.88\times10^{-16}}},
}
where $T(r)$ is the radiation temperature at the horizon reentry $arH=1$, which is related to the temperature at the matter-radiation equality, $T_\eq$, by
\bae{
    (k_\eq r)^{-1}\simeq2(\sqrt{2}-1)\pqty{\frac{g_*}{g_{*\eq}}}^{1/6}\frac{T(r)}{T_\eq},
}
with the horizon scale $k_\eq=\eval{\frac{1}{aH}}_\eq\simeq0.07\Omega_\um h^2$ and the effective degrees of freedom $g_{*\eq}=3.38$ at that time. $\Omega_\um h^2$ is the current matter density parameter.

\begin{figure}
    \centering
    \begin{tabular}{c}
        \begin{minipage}{0.5\hsize}
            \centering
            \includegraphics[width=0.95\hsize]{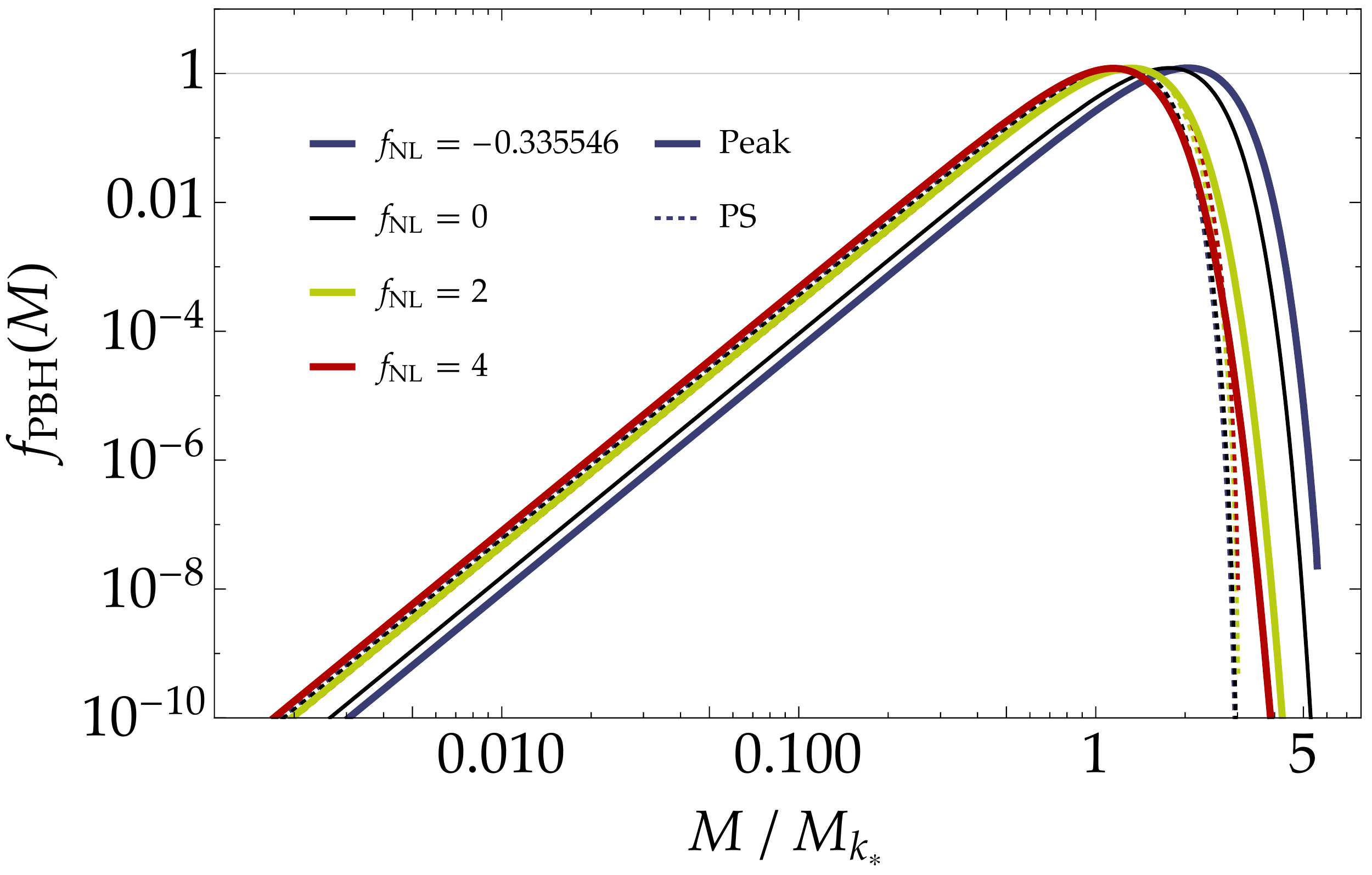}
        \end{minipage}
        \begin{minipage}{0.5\hsize}
            \centering
            \includegraphics[width=0.95\hsize]{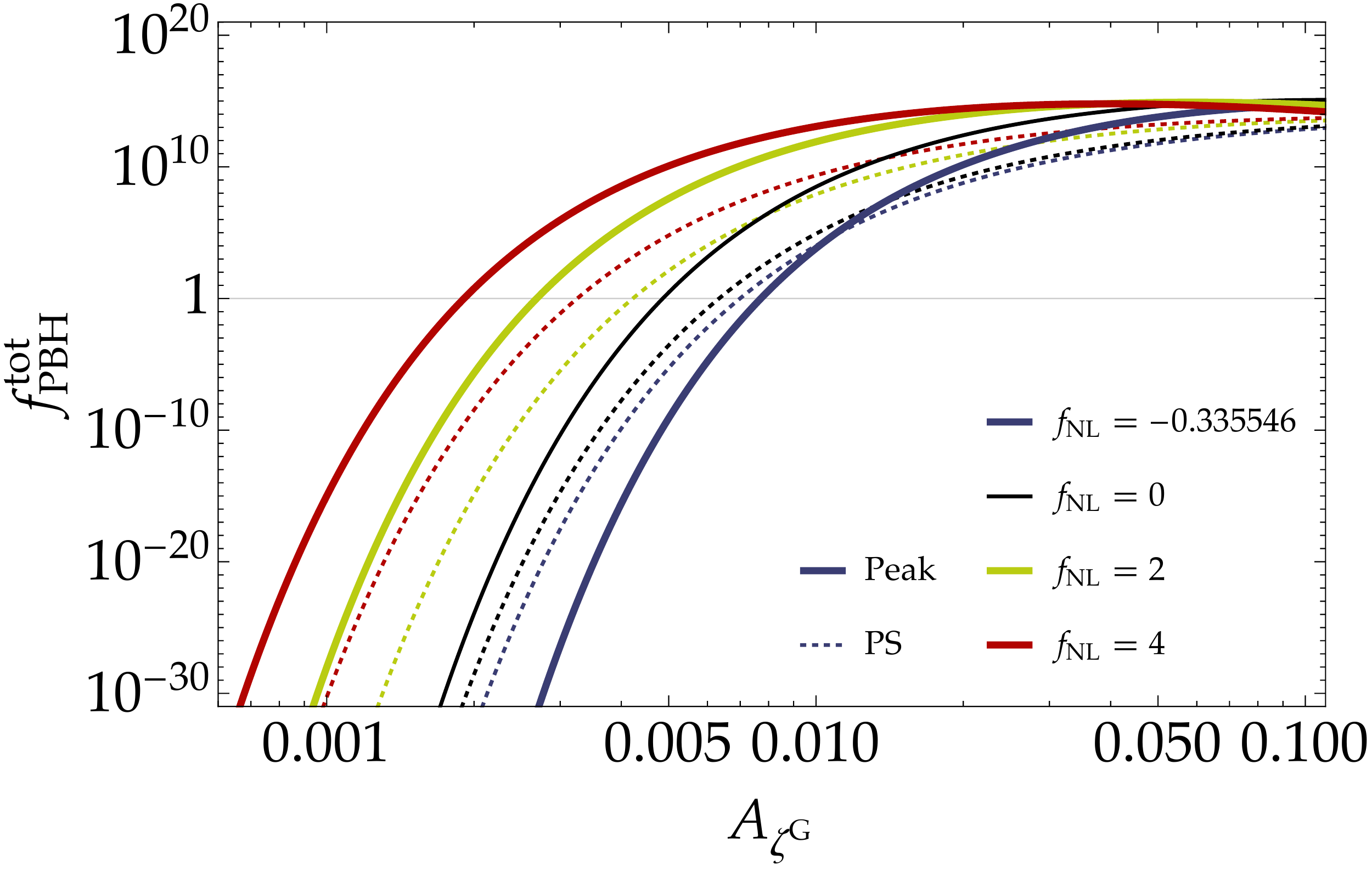}
        \end{minipage}
    \end{tabular}
    \caption{\emph{Left}: PBH mass functions with $\fNL=\fNLmin\simeq-0.336$, $0$, $2$, and $4$ with the same color code to Fig.~\ref{fig: fPBH fNL} in the peak theory (plane) and the Press--Schechter approach (dotted) with tuned amplitudes $A_{\zeta^\uG}$ of the power spectrum such that $f_\PBH^\tot=1$. \emph{Right}: total PBH abundances $f_\PBH^\tot$ as functions of $A_{\zeta^\uG}$ with the same plot style to the left panel.}
    \label{fig: fPBH fNL PS}
\end{figure}

\begin{figure}
    \centering
    \begin{tabular}{c}
        \begin{minipage}{0.5\hsize}
            \centering
            \includegraphics[width=0.95\hsize]{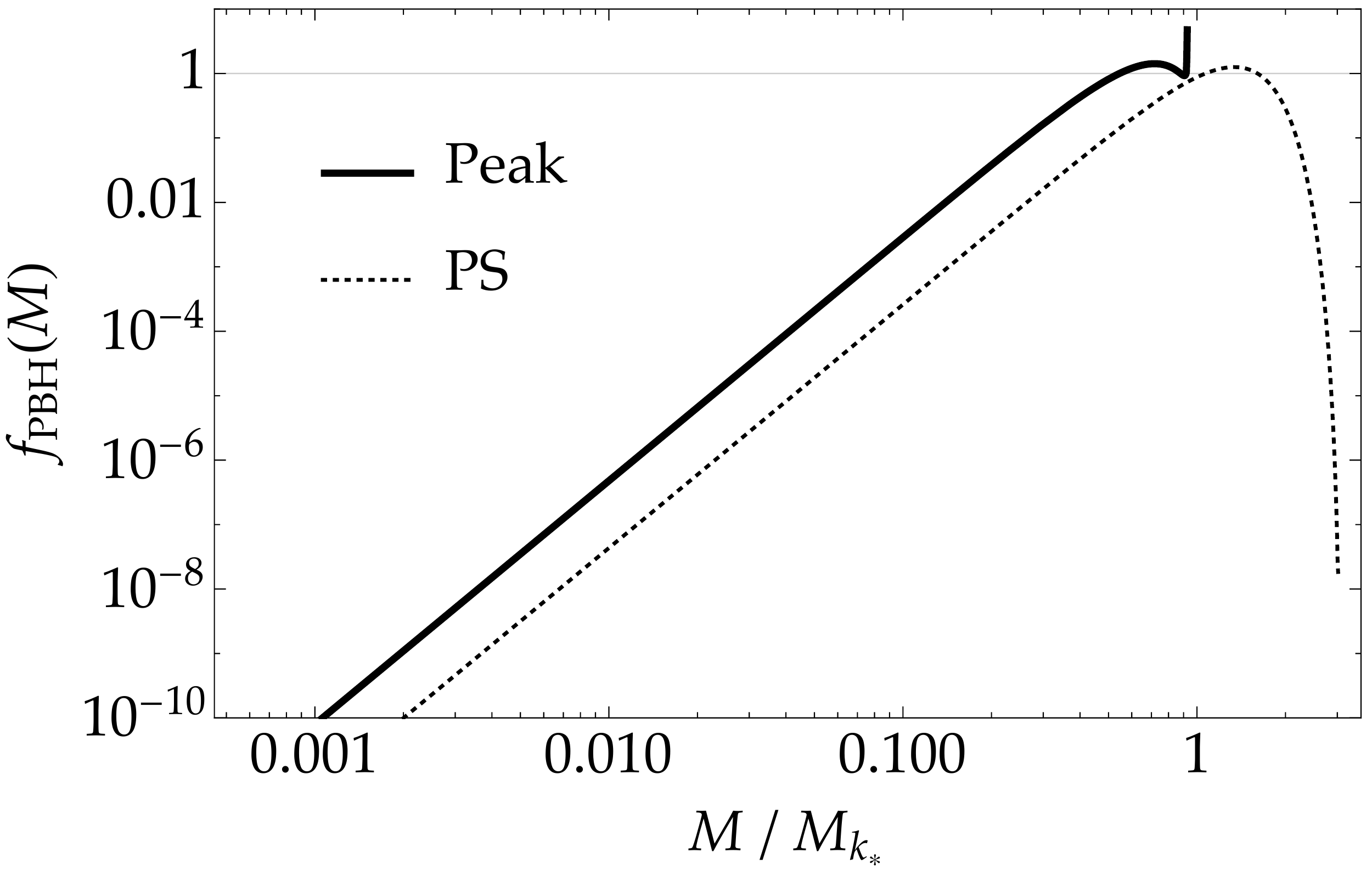}
        \end{minipage}
        \begin{minipage}{0.5\hsize}
            \centering
            \includegraphics[width=0.95\hsize]{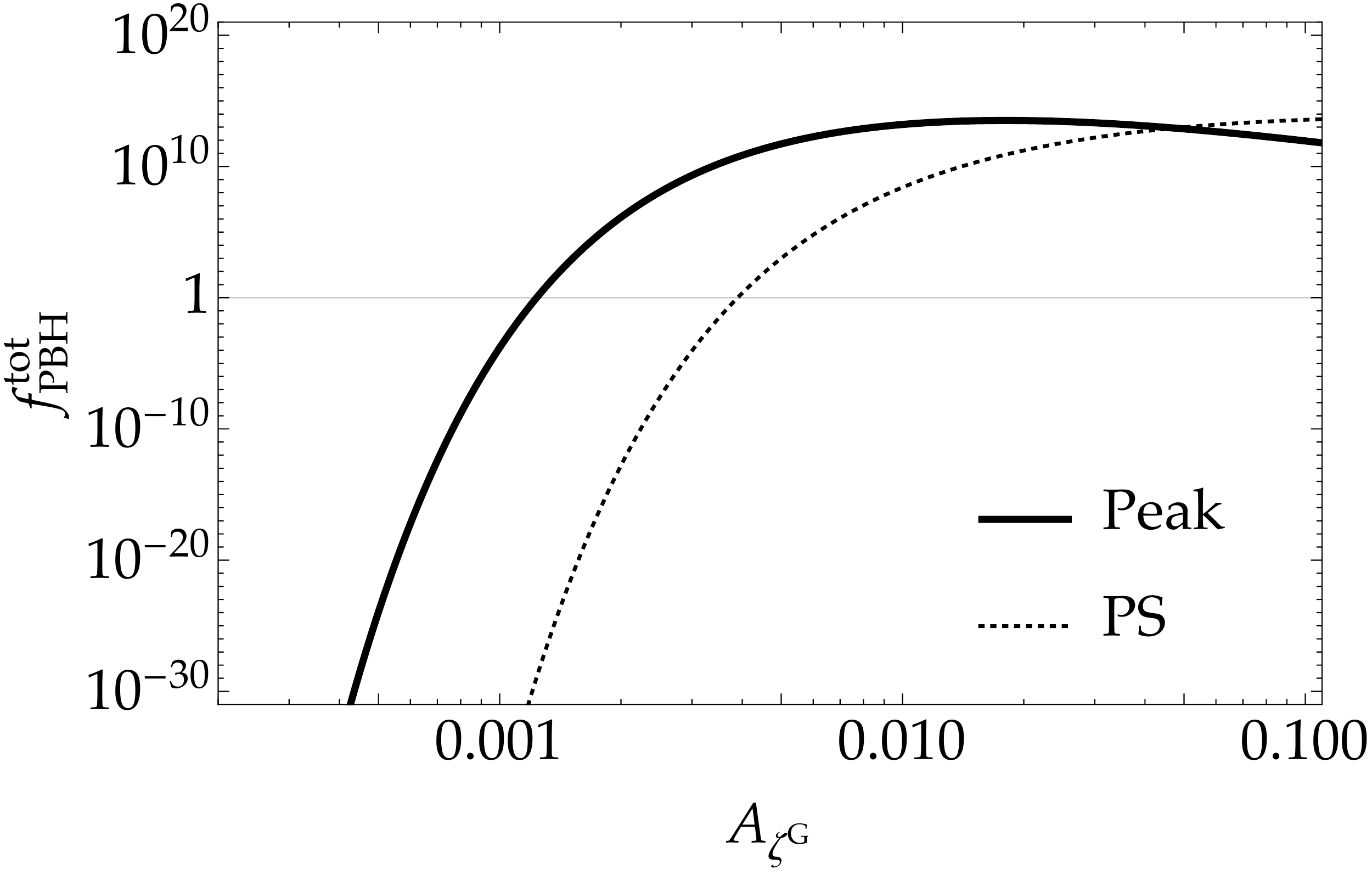}
        \end{minipage}
    \end{tabular}
    \caption{\emph{Left}: PBH mass functions with the exponential tail in the peak theory (thick) and the Press--Schechter approach (dotted) with tuned amplitudes $A_{\zeta^\uG}$ of the power spectrum such that $f_\PBH^\tot=1$. \emph{Right}: total PBH abundances $f_\PBH^\tot$ as functions of $A_{\zeta^\uG}$ with the same plot style to the left panel.}
    \label{fig: fPBH exp PS}
\end{figure}

We show example $f_\PBH$ in the Press--Schechter approach for the $\fNL$ expansion in Fig.~\ref{fig: fPBH fNL PS} and the exponential tail in Fig.~\ref{fig: fPBH exp PS} in comparison to our peak theory.
Their right panels show that the non-Gaussian enhancement/suppression of PBH formation is more significant in the peak theory than the Press--Schechter approach.
The mass spectrum is also insensitive to the non-Gaussianity in the Press--Schechter approach as can be seen in the left panels because it neglects the $\ee^\zeta$ factor in the PBH mass.

\bibliographystyle{JHEP}
\bibliography{main}
\end{document}